\documentclass[11pt,floatfix,prd,preprintnumbers,amssymb]{revtex4}

\usepackage{amsmath}
\usepackage{graphicx}
\usepackage{graphics}
\usepackage{epsfig}
\usepackage{psfrag}
\usepackage{rotating}
\usepackage{dcolumn}
\usepackage{bm}

\newcommand{\beq}{\begin{eqnarray}} 
\newcommand{\eeq}{\end{eqnarray}}

\begin{document}
\baselineskip=16.5pt 
 
\rightline{LPT-ORSAY-09-70} 
\rightline{September 2009}

\vspace{0.7cm} 
 
\begin{center}

{\Large {\bf Higgs boson phenomenology and VEV shift 
\\ in the RS scenario}}
\vspace{0.7cm} 
 
{\large Charles Bouchart, Gr\'egory Moreau}
 
\vspace{0.7cm}

{\it Laboratoire de Physique Th\'eorique, CNRS and Universit\'e Paris--Sud, \\ 
B\^at. 210, F--91405 Orsay Cedex, France.} 
 
\end{center} 
 
\vspace{0.5cm}  
\begin{abstract}   
{\small
In the framework of warped extra dimension models addressing the gauge hierarchy pro\-blem, we consider the Randall-Sundrum (RS) scenario 
under the usual hypothesis of a bulk custodial symmetry. It is shown in detail that there can exist large corrections to the Higgs boson 
Vacuum Expectation Value (VEV) induced by mixings of the gauge bosons with their Kaluza--Klein (KK) excitations. The connection with 
electroweak precision tests is deve\-loped. A noteworthy result is that the correct treatment of the Higgs VEV leads to an increase 
of the lower limit at $95 \% C.L.$ on KK masses that can reach $+30 \%$ (from usually accepted values). 
For a Higgs mass $120 \leq m_h \leq 150$ GeV, the obtained 
limit [from updated precision data] on the first KK gauge boson mass lies in the range $3.3$ - $4.0$ TeV. 
The VEV corrections also play a central role in the corrections to 
the Higgs couplings. We find possibly substantial RS corrections to the various Higgs couplings 
able to affect its phenomenology, starting with a Higgs discovery at LHC more challenging than in the Standard Model (SM). The deviations to
the Higgs production/decay rates found in RS will be testable at ILC as well as 
at LHC. Such RS signatures could even be used at ILC to discriminate among several models beyond the SM. Finally, the possibility of a light Higgs 
boson ($m_h \simeq 99$ GeV) interpreting the excess at $2.3 \sigma$ observed at LEP2 is pointed out.
}
\end{abstract} 
 
\maketitle
 
\newpage


\section{Introduction}
\label{intro}

The Standard Model (SM) represents a successful description of the ElectroWeak (EW) interactions.
Nevertheless, the instability of the EW scale under radiative corrections is a strong indication for a new physics underlying the SM.
The attractive scenario proposed by Randall and Sundrum (RS) \cite{RS} to stabilize the EW scale is based on a 
5--dimensional (5D) theory where the extra dimension is warped and compactified. The non--factorisable metric there is of type Anti--de Sitter
and the space-time, which is thus a slice of $AdS_5$, has two 4--dimensional boundaries: the ultra--violet (UV) boundary at the Planck scale
and the infra--red (IR) brane with an exponentially suppressed scale in the vicinity of the TeV scale. 
The Higgs boson has to be localized at this so--called TeV--brane if the EW scale is to be stabilized by such a geometrical structure.  
\\ In contrast, letting propagate the other SM fields in the bulk \cite{RSloc} allows to suppress higher dimensional operators, potentially 
troublesome with respect to Flavor--Changing Neutral Current effects, by energy scales larger than the TeV scale. This feature has also the
advantage to possibly generate the fermion mass hierarchy and flavor structure by a simple geometrical mechanism \cite{RSloc,RSmass,RSmassBIS}
(even including the small neutrino mass scale \cite{RSnu}). 
In addition, this RS version with bulk matter allows for the unification of gauge couplings at high--energies \cite{UNI-RS}.
More generically speaking, this RS version often turns out to constitute a suitable framework for model building issues.

From a general point of view, those RS models with a fundamental Higgs boson, the alternative models of gauge--Higgs unification \cite{GHunif} and the
Higgsless models \cite{Higgsless} can be thought of as warped extra dimension models constituting dual descriptions, through the $AdS/CFT$ correspondence
\cite{AdSCFT}, of 4D strongly coupled gauge theories (in the limit of a large number of colors) predicting the effective Higgs field 
as a composite state (see e.g. \cite{MHCM}).

Within this new paradigm about the Higgs field sector, one has to take care of the realization of the ElectroWeak Symmetry Breaking (EWSB). 
In particular, additional contributions to the SM observables arise e.g. for the EW gauge boson masses $m_{W,Z}$ and, in turn, for the Fermi coupling constant $G_F$.
More precisely, within the RS scenario, these tree--level $m_{W,Z}$ corrections, with respect to the pure Higgs boson Vacuum Expectation Value (VEV) contribution, originate
from the mixing between zero--mode gauge bosons and their Kaluza--Klein (KK) excitations. 
Even for relatively heavy KK states, this gauge boson mixing, induced by mass mixing terms also driven by
the Higgs VEV, could be quite large since the KK boson wave functions are generally peaked at the TeV--brane where is confined the Higgs boson.
Therefore significant corrections to the SM Higgs VEV can be expected, with various important phenomenological implications.

In the present paper, we first compute the Higgs VEV in the RS context from EW data, including SM loop corrections, 
and find that significant deviations w.r.t. the SM Higgs VEV can be reached in large regions of the allowed parameter space. This computation allows us to
treat precisely the oblique parameters $S$,$T$,$U$ \cite{STU} which depend on the Higgs VEV, and hence to analyze the constraints from precision EW observables.  
A new aspect that matters for these EW Precision Tests (EWPT) is that the correlation between $S$ and $T$ through the RS parameters is modified by the new VEV dependence
on these parameters. These EWPT, based on updated experimental data, are studied under the theoretical assumption of a bulk gauge custodial symmetry
${\rm SU(2)_L\! \times\! SU(2)_R\! \times\! U(1)_X}$ which allows to reduce the final bound on the mass of the first KK gauge boson
excitation $M_{KK}$ (strictly, the KK photon mass) from $\sim 10$ TeV \cite{Burdman} down to roughly a few TeV \cite{ADMS}, 
avoiding then the little hierarchy problem. However, an important result here is that this EWPT analysis leads to more severe bounds on
$M_{KK}$ when the obtained VEV deviations are taken into account: the increase of this limit, at $95 \% C.L.$, can reach 
$\sim +1$ TeV in some extreme situations.

Such increases of the limit on $M_{KK}$ are crucial for the next--coming LHC search of resonances due to the direct production
of KK excitations of gluons or EW gauge bosons \cite{LHCboson} in RS--like models. Those increases of the possible $M_{KK}$ values 
make the most promising tests of KK resonances, which already suffer from low cross sections, even more challenging. Another possibility
for discovering signatures of RS models at next colliders is to observe deviations of the Higgs boson couplings from their values predicted by the SM. 
Such deviations may also of course play a primordial role in the experimental search for the Higgs boson that will probably represents the first goal 
of the LHC.

Hence, we have computed the Higgs boson couplings, 
based on the VEV determination mentioned above. We find that the deviations of the Higgs couplings
from the SM values can be large and also that major parts of those are due to the VEV corrections in RS. The Higgs couplings to pairs of fermions and EW gauge bosons 
as well as the effective couplings at loop level to two gluons or two photons are calculated, taking into account the studied EW precision constraints 
on $M_{KK}$. In particular, we make the connection between the possible RS solution \cite{RSAFB} of the anomaly on the forward--backward
$b$--quark asymmetry $A^b_{FB}$ observed in $e^+e^-$ collisions \cite{AFB-SM} and the cross section $\sigma(g g \to h)$ for the main Higgs production process  
at the LHC: the gluon--gluon fusion mechanism which proceeds through heavy (KK) quark triangular loops. Those past and future collider observables $A^b_{FB}$ and 
$\sigma(g g \to h)$ are indeed connected through the choice of localization in the bulk, for the bottom and top quarks, that fixes their couplings and own masses.                                                      
\\ The obtained reduced production rates at LHC can render the Higgs boson search more difficult in the RS model than in the SM, 
especially during the first phases at low energies and luminosities. Interestingly, the obtained Higgs production rates at LEP2 are also significantly
reduced w.r.t. SM so that the well--known excess at $2.3 \sigma$ in the $Z + b's$ channel can be precisely interpreted by a light Higgs boson ($m_h \simeq  99$ GeV) 
produced within the RS context - thus differently than the way it could also be interpreted within the NMSSM via a reduced $B(h\to b \bar b)$ branching ratio \cite{Gunion}. 
This interpretation is interestingly correlated to the $A^b_{FB}$ solution through the $b$--quark localizations in the bulk.
Such a light Higgs boson lies in a mass range quite difficult to explore at the LHC.   
\\ Concerning the effects in the Higgs boson couplings, the possibly large corrections obtained combined with the expected LHC performance in measurement precisions, 
for the various Higgs production and decay channels, will allow indirect detections/exclusions of RS models in the Higgs sector. 
The future linear collider like ILC will even be capable of precisely measuring the smallest Higgs coupling deviations
arising in RS scenarios, and, discriminate between several models. 
We will illustrate quantitatively those phenomenological results for small Higgs boson masses of $120$ GeV, $150$ GeV
and the entire range of $g_{Z'}$ coupling constant values [associated to the extra $Z'$ boson issued from the enhanced bulk gauge symmetry].

At this stage, one has to mention related works about the Higgs boson couplings within the $AdS/CFT$ paradigm. 
First, the gluon--gluon--Higgs amplitude has been computed in the context of a 5D gauge--Higgs unification as
a dual realization of 4D composite Higgs scenarios \cite{Adam}. $\sigma(g g \to h)$ has also been evaluated in warped extra dimension models \cite{Gautam} 
(based on a bulk custodial symmetry \cite{RSggh,LyonGroup}) without a precise consideration of EWPT. The effective Higgs coupling to two photons has been
compared among various extra dimensional models for some fixed sets of parameters \cite{Gautam,LyonGroup}. Besides, the rate deviations from SM  
predictions, for various Higgs discovery channels at LHC, have been computed within the strongly--interacting light Higgs models \cite{SILH}: the 
LHC will be able to explore values of the cut--off scale $4 \pi f$ (where the model becomes strongly coupled) up to $5-7$ TeV. Finally, the deviations
of the Higgs couplings to EW gauge bosons have been computed in the framework of the so--called 
gaugephobic Higgs boson scenarios \cite{Lillie}. There, the Higgs couplings can be so suppressed that it can be definitely
out of reach for the LHC (in case of a bulk Higgs) \cite{Caccia}.

The organization of the paper is as follows. In Section \ref{VEV}, the corrections to the Higgs VEV arising in the RS framework are computed. 
The connections with EWPT and the Higgs boson phenomenology are studied in the next sections. In Section \ref{EWPT}, an analysis of the
oblique parameters is performed in order to derive the induced bounds on the KK scale. Then in Section \ref{Higgs}, after calculating the Higgs 
couplings, the various tests of the Higgs boson are discussed at the different colliders (LEP, Tevatron, LHC and ILC). Finally, variations on
the main scenario studied are briefly discussed in Section \ref{variants}.

\section{Higgs boson VEV}
\label{VEV}                               

\subsection{Theoretical framework}
\label{formalism}

\noindent {\bf Energy scales:}
Within the RS model,
the gravity scale on the Planck--brane is $M_{\rm Planck}= 2.44\times
10^{18}$ GeV, whereas the effective scale on the TeV--brane $M_{\star}=e^{-\pi
kR_{c}} M_{\rm Planck}$ is suppressed by the warp factor which depends on the curvature
radius of the $AdS$ space $1/k$ and on the compactification radius $R_c$.
For a product $k R_{c} \simeq 11$, $M_{\star}\!=\!{\cal O}(1)$ TeV allowing to
address the gauge hierarchy problem. We will take $k R_{c} \simeq 10.11$ so that
the maximum value of $M_{KK} \simeq 2.45 k e^{-\pi kR_{c}}$, fixed by the theoretical 
consistency bound $k<0.105 M_{\rm Planck}$, is $\sim 10$ TeV in agreement with the
the typical EWPT limits at a few TeV and the range of $M_{KK}$ values considered here.
\\ The parameters noted $c_f$ fix the 5D masses $\pm c_f k$, affected to each fermion $f$, 
and thus control the fermion localizations in the bulk. Those satisfy 
$\vert c_f \vert \! = \! {\cal O}(1)$ to avoid the introduction of new fundamental scales.
\\ \\
\noindent {\bf Gauge symmetry breaking:}
The SM gauge group is recovered after the breaking of the ${\rm SU(2)_R}$ group into ${\rm U(1)_R}$, by boundary conditions
and possibly also by a small breaking of ${\rm SU(2)_R}$ in the bulk effectively parametrized by the 
$\widetilde W^\pm$ mass $\tilde M$ (the $\widetilde W_\mu^\pm$ boson associated to ${\rm SU(2)_R}$ without zero--mode). 
Then the breaking ${\rm U(1)_R \times \rm U(1)_X} \to {\rm U(1)_Y}$ occurs via a VEV on the UV brane:
the state $\widetilde W^3$, associated to ${\rm U(1)_R}$, mixes with $\widetilde B$, associated to ${\rm U(1)_X}$, 
to give the SM hypercharge $B$ boson, the orthogonal linear combination being the extra $Z'$ boson.
The $Z'$ has no zero--mode and its first KK mass is close to $M_{KK}$: $M'_{KK} \simeq 2.40 k e^{-\pi kR_{c}}$.

\subsection{VEV modification}
\label{system}

As in the SM, the Higgs VEV value is determined by the Fermi constant $G_F$. Nevertheless, the corrections to $G_F$
due to the mixing of the SM boson $W^\pm$ with its KK excitations introduce a dependence on the bare EW gauge coupling constants $g$ and $g'$
\footnote{We have checked numerically that the main RS corrections to $G_F$ are oblique, the direct corrections
to the leptonic vertex $Wl\nu$ being negligible in comparison.}.
So within the RS context two other observables must be used to fix the values of the Higgs VEV noted $\tilde v$ and the two bare parameters $g,g'$. 
The natural choice is to use the EW gauge boson masses $m_W$ and $m_Z$. 
The divergent parts of the one--loop EW loop corrections to respectively $G_F$ and $m_{Z,W}$ will affect the $hVV$ coupling ($V=Z^0,W^\pm$), via $\tilde v$, $g$, $g'$, 
but will be canceled out by the SM quantum corrections to the coupling itself: respectively the Higgs boson self--energy and the vertex irreducible correction \cite{SMHiggsCorr}. 
The inputs $G_F$ and $m_Z$ are among the most accurately measured quantities and serve as excellent reference points for EWPT as will be described later. 
The tiny experimental error on $m_W$ (see updated value in \cite{mWEXP}) turns out to not affect significantly the $\tilde v$ value obtained that way.

The $Z$ boson mass defined as the pole of its propagator reads as
\footnote{There is no $\Pi_{3Q}$ neither $\Pi_{QQ}$ terms in the $m_Z^2$ expression as there is no KK mixing involving the photon.},
\begin{equation} 
m_Z^2 = (g^2+g'^2)\frac{\tilde v^2}{4}+(g^2+g'^2) \ \Pi_{33}(m_Z^2) + \delta^{\rm SM}m^2_Z,
\label{MZ2} 
\end{equation}
where $\Pi_{33}(q^2)$ is the vacuum polarization amplitude taking into account the RS--type corrections, more precisely the KK gauge mixing effect \cite{ADMS}:
\begin{equation} 
\Pi_{33}(q^2) = \pi R_{c} \bigg ( \frac{\tilde v^2}{4} \bigg ) ^2 \ \bigg [ (g^2+g'^2)
\ (G^{5D}_{q(++)}-G^{(0)}_{q(++)}) + g_{Z'}^2 \ \cos^4 \theta' \ G^{5D}_{q(-+)} \bigg ].
\label{Pi33} 
\end{equation}
$G^{5D}_{q(++)}$ ($G^{5D}_{q(-+)}$) is the 5D propagator for the $W^3$,$B$ ($Z'$) KK excitations. 
$(++)$ ($(-+)$) indicates Neumann (Dirichlet) and Neumann boundary conditions on the Planck--brane and TeV--brane, respectively.
The massless pole is subtracted from the $W^3$ and $B$ towers.
The new mixing angle is given by $\sin \theta' \equiv \tilde g' / g_{Z'}$ with 
$g_{ Z^\prime}^2 = \tilde g^2 + \tilde g^{\prime 2}$, where $\tilde g$ and 
$\tilde g'$ are respectively the ${\rm SU(2)_ R}$ and ${\rm U(1)_{X}}$ 
couplings; the coupling $g'$ of the SM ${\rm U(1)_Y}$ group reads as $g'=\tilde g 
\tilde g' /g_{Z'}$. One deduces $2 \sin^2 \theta' = 1 \pm \sqrt{1- (2 g'/g_{Z'}) ^2}$.
\\ $\delta^{\rm SM}m^2_Z$ is the SM one--loop correction \cite{SMHiggsCorr}: 
\begin{equation} 
\delta^{\rm SM}m^2_Z = \frac{3 g_Z^2}{16 \pi^2} \sum_{f=t,b} m_f^2 \bigg ( \frac{\mu}{m_f} \bigg ) ^{2\epsilon} 
\ \bigg [ \frac{1}{2\epsilon} + \frac{\pi^2\epsilon}{24} + {\cal O}(\epsilon^2) \bigg ],
\label{DSMZ} 
\end{equation}
$D=4-2\epsilon$ being the space-time dimension in dimensional regularization and $\mu$ the 't Hooft renormalization scale.
For the SM corrections, we restrict to top--quark loops as those constitute the dominant EW corrections to productions and decays of a light Higgs boson ($m_h \ll 2 m_t$). 
Indeed, we will consider Higgs masses below $150$ GeV leading to the lowest EWPT limits on $M_{KK}$ as preferred by little hierarchy arguments.
\\ Similarly, one has,
\begin{equation} 
m_W^2 = g^2 \frac{\tilde v^2}{4} + g^2 \ \Pi_{11}(m_W^2) + \delta^{\rm SM}m^2_W,
\label{MW2} 
\end{equation}
with, 
\begin{equation} 
\Pi_{11}(q^2) = \pi R_{c} \bigg ( \frac{\tilde v^2}{4} \bigg ) ^2 \ \bigg [ g^2 \
(G^{5D}_{q(++)}-G^{(0)}_{q(++)}) + \tilde g^2 \ G^{5D}_{q(-+)} \bigg ],
\label{Pi11} 
\end{equation}
\begin{equation} 
\delta^{\rm SM}m^2_W = \frac{3 g^2}{16 \pi^2} \bigg ( \frac{\mu}{m_t} \bigg ) ^{2\epsilon} 
\ \bigg [ \frac{m_b^2+m_t^2}{2\epsilon} + \frac{m_b^2+m_t^2}{4} + \frac{m_b^4 \ ln(m_b^2 / m_t^2)}{2(m_t^2-m_b^2)} + {\cal O}(\epsilon) \bigg ].
\label{DSMW} 
\end{equation}
Taking the limit $q^2 \to 0$, one obtains
\begin{equation} 
\frac{1}{4\sqrt{2}G_F} = \frac{\tilde v^2}{4} + \Pi_{11}(0) + \delta^{\rm SM}G_F^{-1},
\label{GF} 
\end{equation}
with at first order in $1 / k \pi R_{c}$ 
\begin{equation} 
\Pi_{11}(0) = - k \pi R_{c} \frac{g^2}{32} \bigg ( \frac{\tilde v^2}{ke^{-\pi kR_c}} \bigg )^2 
\ \bigg [ 1-\frac{1}{k \pi R_{c}} \bigg ]
- k \pi R_{c} \frac{\tilde g^2}{32} \bigg ( \frac{\tilde v^2}{ke^{-\pi kR_c}} \bigg )^2 
\ \bigg [ 1-\frac{\tilde M^2}{4 k^2} \bigg ],
\label{Pi110} 
\end{equation}
and,
\begin{equation} 
\delta^{\rm SM}G_F^{-1} = \frac{6}{(8 \pi)^2} 
\bigg [ \frac{m_b^2+m_t^2}{2} + \frac{m_b^2 \ ln(m_b^2 / m_t^2)}{(1-m_b^2/m_t^2)} \bigg ].
\label{DSMG} 
\end{equation}

Numerically, the SM loop corrections are small compared to the RS tree--level ones, as we are going to see.
We do not consider higher order corrections involving both loops and KK excitations.
Hence, the quantities $\tilde v$, $g$ and $g'$ can be determined within given RS scenarios 
by solving the system formed by the three equations: Eq.(\ref{MZ2}), Eq.(\ref{MW2}) and Eq.(\ref{GF}).
We have checked that the computation of the gauge boson masses through Eq.(\ref{MZ2})-(\ref{MW2}) is equivalent to their
calculation within the perturbation approach in the limit of the inclusion of a large number of KK excitations [see the description
of the KK gauge boson mass matrix in Section \ref{couplings}].
The obtained value for the coupling constant $g$ is $g=0.653$ in both the SM and RS models, which means with and without the corrections due to KK mixing. 
$g'$ takes similar values in the SM and RS models: $g'=0.356$ and $g'=0.357$ respectively.
The numerical results for $\tilde v$ with characteristic values of the parameters $M_{KK}$ and $g_{Z'}$ are shown in Table \ref{MainTable}.
The motivation for the chosen values of $M_{KK}$ will become clear in Section \ref{EWPT}. 
The allowed range for the $g_{Z'}$ value is defined as follows. 
The minimum $g_{Z'}$ value is equal to $2 g' \simeq 0.72$ for consistency reasons about the $\theta'$ mixing angle.
On the other side, the perturbativity condition for the $Z'$ boson coupling is that $2 k \pi R_c g_{Z'}^2 Q_{Z'}^2 / 16 \pi^2$ must be smaller than unity
according to the naive dimensional argument \cite{MHCM,LHCboson}. The reason being that the effective 4D coupling of $Z'$ is increased by a factor 
as large as $\sqrt{2 \pi k R_c}$ for fields at the TeV--brane. In general even for reproducing the correct top mass, the top quark field is not exactly
located on the TeV--brane but has only a peaked profile there, so that the real overlap factor should be significantly smaller than $\sqrt{2 k \pi R_c}$.
We will consider the coupling constant constraint $g_{Z'} < 2 \sqrt{2\pi / k R_c} \simeq 1.57$, 
keeping in mind that the new charge $Q_{Z'}$ is an additional source for suppressing the $Z'$ coupling
given e.g. the ${\rm SU(2)_R}$ fermionic representations addressing the $A^b_{FB}$ anomaly \cite{RSrep}.
\\ The first observation from Table \ref{MainTable}
is that the RS induced deviations to the Higgs VEV are large compared to the pure SM quantum corrections: 
the Higgs VEV is at the reference value of $v=246$ GeV in the SM at tree--level
and at $v_{SM}=245$ GeV after including the one--loop EW corrections in Eq.(\ref{GF}) for $m_t=173.1$ GeV \cite{mtopEXP}.
Because of the negative sign of $\Pi_{11}(0)$, the variation of the Higgs VEV induced 
in RS is positive as show the $\tilde v$ values in Table \ref{MainTable} which are significantly
larger than $v_{SM}$. 
This increase of the VEV is higher with the reduction of $M_{KK}$ (e.g. at fixed $g_{Z'}=1.57$) 
or with the enhancement of $g_{Z'}$ is explained by a larger KK gauge mixing effect
in Eq.(\ref{MZ2})-(\ref{MW2})-(\ref{GF}). The increase of the KK mixing with $g_{Z'}$ will be clear from the texture of the KK gauge boson mass matrix
(see Appendix \ref{MassMat}).

\begin{table}[!ht]
\begin{center}
\begin{tabular}{|c|c|c|}
\hline  & & \\
{\bf A]} $ \ m_h = 120$ GeV, $g_{Z'} = 1.57 \  \ $  & 
{\bf B]} $ \ m_h = 120$ GeV, $g_{Z'} = 0.72 \  \ $  &  
{\bf C]} $ \ m_h = 150$ GeV, $g_{Z'} = 1.57 \  \ $    
\\ 
& & 
\\ \hline & &   \\
$M_{KK} = 4025$ GeV                  & $M_{KK} = 3370$ GeV                  &  $M_{KK} = 4095$ GeV     
\\ 
& &
\\ \hline  & & \\
$\tilde v = 322$ GeV                 & $\tilde v = 257$ GeV                 &  $\tilde v = 311$ GeV     
\\ & & \\ \hline  & & \\
$g^{RS}_{hZZ}/g^{SM}_{hZZ}  = 57.3\%$   &  $ g^{RS}_{hZZ}/g^{SM}_{hZZ}   = 87.2\%$ &  $ g^{RS}_{hZZ}/g^{SM}_{hZZ} = 60.1\%$
\\  & & \\ \hline  & & \\ 
$g^{RS}_{hWW}/g^{SM}_{hWW} = 57.5\%$   & $g^{RS}_{hWW}/g^{SM}_{hWW} = 87.4\%$  &  $g^{RS}_{hWW}/g^{SM}_{hWW} = 60.3\%$
\\  & & \\ \hline  & & \\ 
$\lambda^{RS}_{\tau}/\lambda^{SM}_{\tau} = 76.2\%$   & $\lambda^{RS}_{\tau}/\lambda^{SM}_{\tau} = 95.5\%$  &  $\lambda^{RS}_{\tau}/\lambda^{SM}_{\tau} = 79.1\%$
\\  & & \\ \hline  & & \\ 
$\lambda^{RS}_{b}/\lambda^{SM}_{b} = [71 , 75]\%$   & $\lambda^{RS}_{b}/\lambda^{SM}_{b} = [90 , 93] \%$ & $\lambda^{RS}_{b}/\lambda^{SM}_{b} = [74 , 78]\%$
\\  & & \\ \hline  & & \\ 
  $g^{RS}_{hgg}/g^{SM}_{hgg} = [77.6 , 80.8]\%$   
& $g^{RS}_{hgg}/g^{SM}_{hgg} = [96.2 , 99.1]\%$  
& $g^{RS}_{hgg}/g^{SM}_{hgg} = [80.1 , 83.3]\%$
\\  & & \\ \hline  & & \\ 
  $g^{RS}_{h\gamma\gamma}/g^{SM}_{h\gamma\gamma} = [74.9 , 75.1]\%$   
& $g^{RS}_{h\gamma\gamma}/g^{SM}_{h\gamma\gamma} = [100.6 , 100.8]\%$
& $g^{RS}_{h\gamma\gamma}/g^{SM}_{h\gamma\gamma} = [77.1 , 77.3]\%$
\\  & & \\ \hline
\end{tabular}
\end{center}
\caption{Numerical results for the lower limit at $95.45 \% C.L.$ on $M_{KK}$ (from EWPT) and for the Higgs boson VEV within the RS scenario, 
with three characteristic sets of parameters [called A, B and C]. $\tilde M/k = 0.11$, $0.41$ and $0.12$ respectively (optimized values leading 
to the minimum $M_{KK}$ limits) for the points A, B and C. Note that
$\tilde v$ is obtained for the indicated allowed $M_{KK}$ value, and reciprocally, the $M_{KK}$ limit corresponds to the given Higgs VEV.
For these 3 points, the values for the ratios (RS over SM) of the effective Higgs couplings to EW gauge bosons ($Z$, $W$, $\gamma$), 
gluons ($g$) and fermions ($\tau$, $b$) are also given. The Yukawa coupling ratios for the other light fermions are identical 
to the $\tau$ lepton one given here.}
\label{MainTable}
\end{table}

\section{EW precision tests}
\label{EWPT}

The mixings with KK gauge boson excitations induce modifications of the EW gauge boson propagators, the oblique corrections, 
that can be parametrized by the three $S_{\rm RS}$,$T_{\rm RS}$,$U_{\rm RS}$ quantities \cite{STU}
(we define those such that they vanish in the absence of KK mixing). 
Here, we are interested in variations of the constraints from considerations on $S$,$T$ due to the corrections on the Higgs VEV.
There exist also constraints on the $W,Y$ parameters \cite{BPRS}
\footnote{The present model does not belong to the category of universal theories considered in \cite{BPRS}.}. Compared to $S$,$T$,
these parameters are proportional to higher derivatives of the vacuum polarization amplitudes, involved in higher order terms of the $\Pi_V(q^2)$ 
expansion, that are not considered within the present RS models.
$W$ and $Y$ allow to take into account the relevant LEP2 observables, namely, the differential cross sections for $e^+e^- \to f \bar f$ 
which we already partially include in this analysis: the process $e^+e^- \to b \bar b$ is included in the $A^b_{FB}$ solution above the $Z$ pole 
studied in Section \ref{couplings}.
The corrections to EW observables in the third quark generation sector [$b$ and $t$] are treated separately through a fit  
independent from the oblique parameters.
In contrast, for the light SM fermions, the associated parameters $c_{\rm light}$ are taken larger than $0.5$
to generate small masses \cite{RSmass} and to minimize their couplings to KK gauge boson excitations (and thus corrections 
to EW observables). Then the fermion--Higgs higher--dimensional operators, obtained after having integrated  
out heavy KK modes, get a special form which allows one to redefine their effects into purely oblique corrections \cite{ADMS}. 
The effects from the effective 4--fermion operators are negligible \cite{ADMS} for $c_{\rm light}>0.5$ and $M_{KK} \geq 3$ TeV.

For $c_{\rm light} \geq 0.5$, the oblique parameter $S_{\rm RS}$ reads as (not writing terms suppressed e.g. by $1 / k \pi R_c$ factors \cite{ADMS,DavSonPer}
which are however included numerically),
\begin{equation}   
S_{\rm RS} \simeq 2 \pi \bigg ( \frac{2.45 \ \tilde v}{M_{KK}} \bigg ) ^2 
-
\pi^2 k R_c \ \frac{g^2 + g^{\prime 2} + g_{Z'}^2 \cos^4 \theta'}{16} \bigg ( \frac{2.45 \ \tilde v}{M_{KK}} \bigg ) ^4 ,
\label{SRS}   
\end{equation}
where the first term is the contribution from fermion--Higgs higher--dimensional operators. 
The other term comes from the gauge--Higgs sector and is at order $(\tilde v/M_{KK})^4$ thus smaller than the first one. 
Here, we insist on the fact that $g$ and $g'$ are the bare coupling constants, that are computed as described in Section \ref{system}.
It was shown recently that the KK mixing induces one--loop contributions (considering only the Higgs sector \cite{BurdmanBIS}) to the $S$ parameter 
which are not finite and are cut--off dependent (or depend on an energy scale). The physical $m_h$ dependence in $S$ should be approached with care in such a 5D theory as  
it can be affected by a renormalization procedure. 
Therefore here, given our goal of studying the Higgs VEV correction effect, we do not consider higher order corrections to $S$ involving both loops and KK excitations. 
\\ the oblique parameter $T_{\rm RS}$ is given by,
\begin{equation}   
\alpha_0 T_{\rm RS} 
\equiv \frac{g^2+g'^2}{m_Z^2} (\Pi_{11}(0)-\Pi_{33}(0))
\ \simeq k \pi R_c \ \frac{g^2+g'^2}{m_Z^2} \ \frac{\tilde g^2}{32} \ \frac{\tilde M^2}{4k^2} \ \bigg ( \frac{2.45 \ \tilde v^2}{M_{KK}} \bigg ) ^2 ,
\label{TRS}   
\end{equation}
where $\alpha_0 = \alpha (q^2=0)$ is the QED fine structure constant given e.g. in \cite{PDG} and
$\tilde M$ is the $\widetilde W^\pm$ mass originating from the small bulk breaking of ${\rm SU(2)_R}$. 
This expression has the interest of illustrating the custodial protection of $T_{\rm RS}$. 
Other scenarios, including in particular the case where ${\rm SU(2)_R}$ remains unbroken in the bulk, are discussed later. 
\\ The parameter $U_{\rm RS}$ is non--vanishing only at the order $(\tilde v/M_{KK})^4$ in the KK expansion, in contrast with $S_{\rm RS}$ and $T_{\rm RS}$. 
Hence, for the relevant values of RS parameters, the $U_{\rm RS}$ values obtained are totally negligible compared to $S_{\rm RS}$, $T_{\rm RS}$. 
We thus fix $U_{\rm RS}$ at zero in an extremely good approximation.

The corrections to EW observables measured up to the $m_Z$ scale can be expressed in function of the three variables $S_{\rm RS}$,$T_{\rm RS}$,$U_{\rm RS}$
(we keep $U_{\rm RS}$ at this stage for the completeness of given formulas).
We concentrate on the experimental measurements of $m_W$, $\sin^2\theta_{{\rm eff}}^{{\rm lept}}$ 
\footnote{$\theta_{{\rm eff}}^{{\rm lept}}$ denotes the {\it Weinberg angle} modified by radiative corrections 
and the exponent ``lept'' means that it is the value which can be obtained directly from lepton asymmetry measurements.}
and the partial $Z$ width into charged leptons $\Gamma_{\ell \ell}$ 
as those are the most precise and crucial in constraining the plan $\{ T,S \}$ \cite{PDG}.
The theoretical expression for the observable $m_{W^\pm}$ reads as
\begin{equation}   
m_W^2 = m_W^2 \vert_{ref} + \frac{\alpha_0 \ c^2}{c^2-s^2} m_Z^2 \left ( - \frac{1}{2} \ S_{\rm RS} + c^2 \ T_{\rm RS} + \frac{c^2-s^2}{4 s^2} \ U_{\rm RS} \right )  
\label{mWexp}   
\end{equation}
where $m_{W^\pm} \vert_{ref}$ represents the value calculated as accurately as possible within the pure SM
and $s$ ($c$) stands for $\sin \theta_W$ ($\cos \theta_W$), $\theta_W$ being the EW mixing angle.
Here we emphasize that the strict expression for $m_W$ involves the bare value of the EW mixing angle $\theta_W$. 
The expression for $\sin^2\theta_{{\rm eff}}^{{\rm lept}}$ is
\begin{equation}   
\sin^2\theta_{{\rm eff}}^{{\rm lept}} = 
\sin^2\theta \vert_{ref} + \frac{\alpha_0}{c^2-s^2} \left ( \frac{1}{4} \ S_{\rm RS} - s^2 c^2 \ T_{\rm RS} \right )  .
\label{sinexp}   
\end{equation}
The partial $Z$ width reads as,
\begin{equation}   
\Gamma_{\ell \ell} = 
\Gamma_{\ell \ell} \vert_{ref} + \alpha_0 \ \Gamma_{\ell \ell}^0 \bigg (
- \frac{2(1-4 \sin^2 \theta_0)}{(1+(1-4 \sin^2 \theta_0)^2)} \frac{S_{\rm RS}}{c^2-s^2} 
+ \bigg \{ 1 + \frac{8(1-4 \sin^2 \theta_0)}{(1+(1-4 \sin^2 \theta_0)^2)} \frac{s^2c^2}{c^2-s^2} \bigg \} T_{\rm RS}
\bigg ) ,
\label{Gllexp}   
\end{equation}
with, 
$$
\Gamma_{\ell \ell}^0 = \frac{m_Z^3 \ G_F}{3\sqrt{2}\pi} (2\sin^4 \theta_0-\sin^2 \theta_0+\frac{1}{4}).
$$
$\sin \theta_0$ involves the electroweak mixing angle obtained in the improved Born approximation, namely by taking into account only 
the well--known QED running of $\alpha$ up to $m_Z$: 
$$
\sin^2 \theta_0 = \frac{1}{2} \bigg [ 1 - \sqrt{ 1 - 4 \frac{\pi \alpha}{\sqrt{2} G_F m_Z^2} } \bigg ] = 0.2310
$$
where $\alpha = \alpha (q^2=m_Z^2) = \alpha_0/(1 - \Delta \alpha)$ \cite{HiggsReviewI} with 
$\Delta \alpha = \Delta \alpha_{lept} + \Delta \alpha_{had} + \Delta \alpha_{top}$,
$\Delta \alpha_{lept}=0.0315$ \cite{alpha} and $\Delta \alpha_{top} = - 0.00007$ \cite{alpha}.
By virtue of the theorem obtained in Ref.~\cite{thm}, only SM fermions contribute to the $\alpha$ running and hence to the leptonic
contribution $\Delta \alpha_{lept}$, the hadronic contribution $\Delta \alpha_{had}$ and the top quark one $\Delta \alpha_{top}$.
\\ At the moment, there is a small controversy on the precise experimental value
of $\Delta \alpha_{had}(m_Z^2)$. 
This value enters in our EWPT analysis through $\sin^2 \theta_0$ as well as through the SM loop calculations of the
fitted observables (see later). 
Recent measurements of $\Delta \alpha_{had}$ \cite{PDG} via the $e^+e^-$ annihilation cross sections give the results 
$0.02758 \pm 0.00035$ \cite{Daee58}; $0.027594 \pm 0.000219$ \cite{Daee59}; $0.02768 \pm 0.00022$ \cite{Daee68}, 
whereas the measurement gives $\Delta \alpha_{had}=0.02782 \pm 0.00016$ using the hadronic $\tau$ decays \cite{Datau82}.
The consensus is still to use the $e^+e^-$ data, which are close to each other. We thus use $\Delta \alpha_{had} = 0.02768 \pm 0.00022$,
keeping in mind that the result from $\tau$ data is higher. 
If $\tau$ data are used in the calculation of $a_\mu^{SM}$, the muon magnetic anomaly discrepancy decreases 
(from $3.1 \sigma$ for $e^+e^-$-based results \cite{DavierLAST}) down to $1.8 \sigma$ (see recent updates in \cite{Datau82,DavierLASTBIS}).
While almost solving $\Delta a_\mu$, the $\tau$ data, which raise the value of $\Delta \alpha_{had}$, lead to an EWPT upper bound
on $m_h$ of $133$ GeV \cite{Datau82} leaving a narrow window for the SM Higgs mass, given the direct LEP2 lower bound of $114.4$ GeV \cite{DirectMH}.
It is interesting to note that such a tension does not appear, within the RS framework, where both the EWPT limit can be higher
(there is possibly a positive contribution to $T$) and the LEP2 limit smaller [see later]. For example, with $M_{KK}=4.1$ TeV, 
$g_{Z'}=1.51$ and $\tilde M/k = 0.14$, the EWPT limit at $95 \% C.L.$ on $m_h$ is $150$ GeV for $\Delta \alpha_{had}=0.02782$.

\begin{figure}[!ht]
\begin{center}
\includegraphics[width=12.cm]{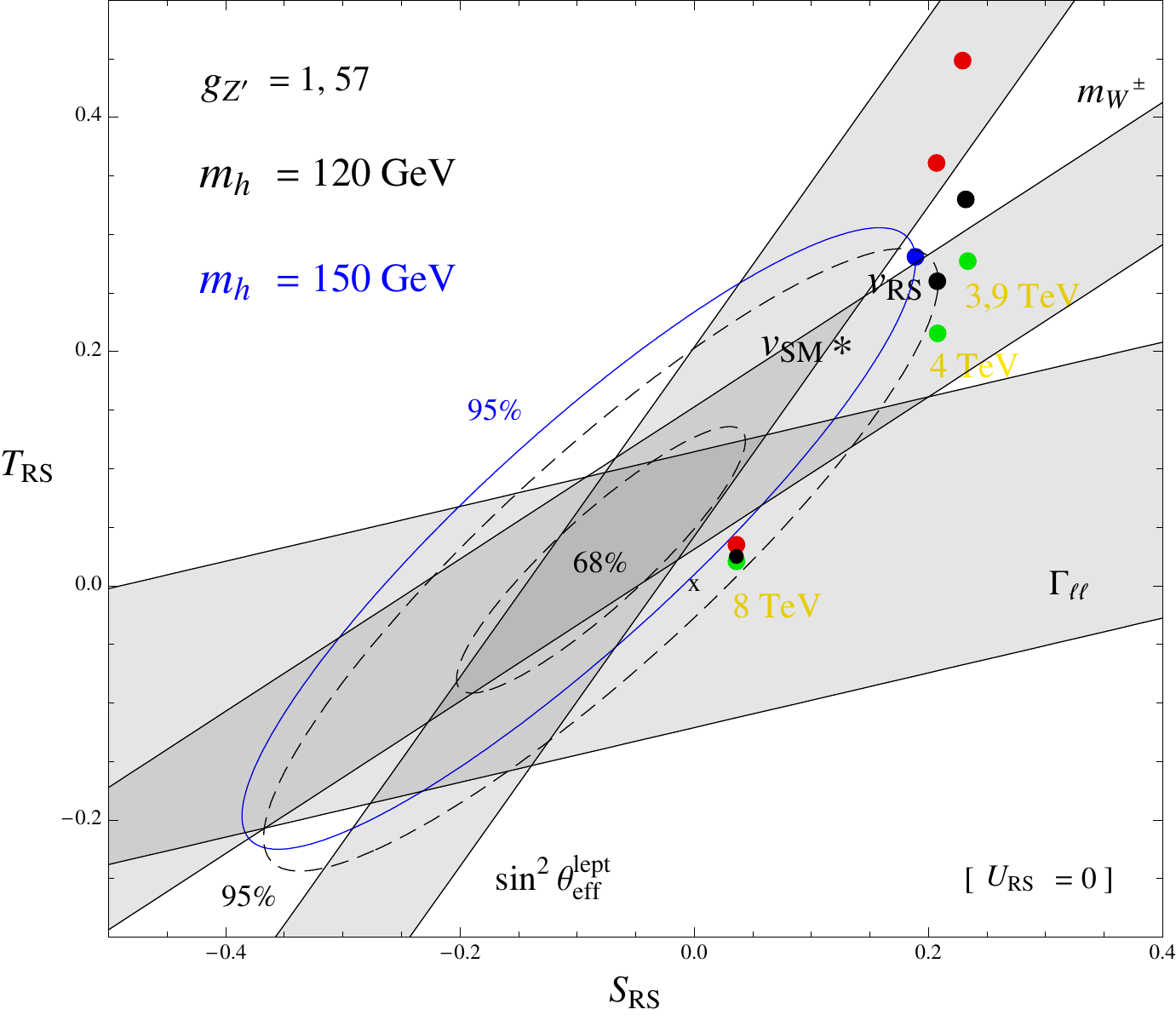} 
\end{center}
\vspace*{-5mm}
\caption{Contour levels in the plan $\{ T_{\rm RS},S_{\rm RS} \}$ at $68.27 \% C.L.$ and $95.45 \% C.L.$ for the fit of $m_{W}$, 
$\Gamma_{\ell \ell}$ and $\sin^2\theta_{{\rm eff}}^{{\rm lept}}$: the two ellipses in dashed--line are for $m_h = 120$ GeV, 
whereas the blue one in plain--line is for $m_h = 150$ GeV. 
Also shown, for $m_h = 120$ GeV, are the $1\sigma$ bands for the experimental values of the considered EW observables.
We use the exact results from the two--loop calculations for $m_{W} \vert_{ref}$ 
(valid for $100$ GeV $ \lesssim m_h< 1$ TeV) \cite{MwSM}, 
$\Gamma_{\ell \ell} \vert_{ref}$ ($75$ GeV $<m_h< 350$ GeV) \cite{GllSM} and $\sin^2\theta \vert_{ref}$ 
($10$ GeV $<m_h< 1$ TeV) \cite{sinSM}. 
The other quantities, on which these SM predictions depend, are fixed at $m_t=173.1 \pm 1.3$ GeV (in agreement with the recent 
Tevatron data \cite{mtopEXP}), $\alpha_s=0.1204 \pm 0.0009$ \cite{alpha,LEPex,Seb}.
The plot is obtained for $g_{Z'}=1.57$ and $U_{\rm RS} \simeq 0$.
The three red, black, green points are associated to the theoretical predictions for the coordinates $S_{\rm RS}$,$T_{\rm RS}$ 
with $\tilde M/k=0.13$; $0.11$; $0.10$ respectively. Those correspond to $m_h = 120$ GeV 
and $M_{KK}=3980$, $4025$, $8000$ GeV from right to left -- as indicated.
While the blue point is for $m_h = 150$ GeV, $M_{KK}=4095$ GeV and $\tilde M/k=0.12$, 
the star corresponds to $m_h = 120$ GeV, $M_{KK}=4025$ GeV and $\tilde M/k=0.11$ 
but using $v_{SM}=245$ GeV (rather than $v_{RS}= \tilde v = 322$ GeV). The cross is located at the origin.}
\label{fig:ST}
\end{figure}

As shows e.g. Eq.(\ref{sinexp}), the accurate measurements of EW observables translate into limits in the plan $T_{\rm RS}$ versus $S_{\rm RS}$.  
These limits depend on the SM expectation e.g. noted $\sin^2\theta \vert_{ref}$ for $\sin^2\theta_{{\rm eff}}^{{\rm lept}}$. In general, the precise 
predictions of EW observable values calculated within the SM from QCD/EW corrections \cite{MwSM,GllSM,sinSM}
depend in turn on the top and Higgs masses as well as on the strong coupling constant and the photon vacuum polarization $\Delta \alpha$.
\\ In Fig.(\ref{fig:ST}) are presented the limits in the plan $\{ T_{\rm RS},S_{\rm RS} \}$ [for $m_h = 120$ GeV] corresponding to values of $m_W = 80.399 \pm 0.025$ GeV 
(combined LEP2 and Tevatron Run II data) \cite{mWEXP}, 
$\Gamma_{\ell \ell} = 83.985 \pm 0.086$ MeV (single lepton channel) \cite{alpha,LEPex} and $\sin^2\theta_{{\rm eff}}^{{\rm lept}}$ within
$1 \sigma$ deviation from their experimental central value. The experimental value used here for $\sin^2\theta_{{\rm eff}}^{{\rm lept}}$
is a combination of the 5 values resulting from the 5 asymmetry measurements: $A^\ell_{FB}(m_{Z})$, ${\cal A}_\ell (P_\tau)$,  
${\cal A}_\ell (SLD)$, $A^c_{FB}(m_{Z})$ and $Q^{had}_{FB}$ \cite{alpha,LEPex}.
\\ In Fig.(\ref{fig:ST}) are also shown the contour levels in $\{ T_{\rm RS},S_{\rm RS} \}$ [$m_h = 120$; $150$ GeV] 
associated to the confidence levels at $68.27 \% C.L.$ and $95.45 \% C.L.$. 
Those result from a $\chi^2$--analysis of the fit between the RS theoretical predictions for the considered observables and their respective experimental value.  
In particular, the $95.45 \%$ level corresponds to $\chi^2 / d.o.f.= 12.85 / 6$.

The correlation between $S_{\rm RS}$ and $T_{\rm RS}$ through $M_{KK}$, $g_{Z'}$ is modified by the new dependence of the Higgs VEV ($\tilde v$ entering Eq.(\ref{SRS})-(\ref{TRS})) 
on these parameters and on $\tilde M/k$ (via Eq.(\ref{MZ2})-(\ref{Pi110})). 
In particular, the lower limit at $95.45 \% C.L.$, from the global EW fit, on $M_{KK}$ as a function of $\tilde M/k$ possesses a minimum depending on $g_{Z'}$ and $m_h$. 
This feature is illustrated by the fact that, if e.g. $g_{Z'}=1.57$ and $m_h=120$ GeV, the lower limit is $M_{KK}=4025$ GeV for $\tilde M/k = 0.11$ 
\footnote{The $\tilde M$ values considered are at worst an order of magnitude close to $k$, so that no new energy scale is introduced in the RS scenario, 
but remain smaller than $k$ so that a small bulk breaking of the custodial--isospin is guaranteed.} (theoretical point in 
black exactly on the $95.45 \% C.L.$ contour in the plan $\{ T_{\rm RS},S_{\rm RS} \}$ of Fig.(\ref{fig:ST})) whereas it is higher than $4025$ GeV for $\tilde M/k = 0.10$ (green point
on the $95.45 \% C.L.$ contour for $8000>M_{KK}>4025$ GeV) as well as $\tilde M/k = 0.13$ (red points).
The lower limit on $M_{KK}$ at $95.45 \% C.L.$ (for an optimized $\tilde M/k$ value corresponding to the minimum of that limit) increases with $\tilde v$ (making $S_{\rm RS}$
and $T_{\rm RS}$ larger) and hence with $g_{Z'}$, as shows Table \ref{MainTable} for $m_h=120$ GeV. Finally, the effect of the increase of $m_h$ on the $M_{KK}$ limit is an
enhancement (as shows Table \ref{MainTable}) as it shifts the $95.45 \% C.L.$ ellipsoidal contour according to Fig.(\ref{fig:ST}), so that a larger $M_{KK}$ is required to
decrease $\tilde v$ and directly the $S_{\rm RS}$ coordinate of the theoretical point. Fig.(\ref{fig:ST}) also illustrates that the behavior of the theoretical
predictions for $S_{\rm RS}$ and $T_{\rm RS}$ in the limit of high $M_{KK}$ is a convergence to zero. By consequence, for too large $M_{KK}$'s in the case e.g. $m_h=150$ GeV, 
the theoretical points go out of the $95.45 \% C.L.$ ellipse as this one does not contain the origin. There exist thus an upper limit at $95.45 \% C.L.$ on $M_{KK}$ due 
to EWPT which is $M_{KK}<4490$ GeV for $g_{Z'}=1.57$, $\tilde M/k = 0.12$ (even at this upper limit $\tilde v =286$ GeV i.e. there is still a significant RS contribution).

An important result is the impact of the Higgs VEV corrections on the EWPT constraint on $M_{KK}$.
Let us consider for instance the point A of parameter space given in Table \ref{MainTable} and represented in Fig.(\ref{fig:ST}) as the black point sitting on the $95.45 \% C.L.$ limit: 
replacing now the `should--be' Higgs VEV $\tilde v = 322$ GeV by the SM value $v_{SM}=245$ GeV,   
both the $S_{\rm RS}$ and $T_{\rm RS}$ values are reduced and the black point becomes the black star on the plot. 
To shift this star back to the $95.45 \% C.L.$ ellipsoidal contour, one can decrease $M_{KK}$
(at fixed $g_{Z'}$ and $m_h$) which would increase $S_{\rm RS}$ and $T_{\rm RS}$. Indeed, for $g_{Z'}=1.57$, $m_h=120$ GeV and the optimized $\tilde M/k = 0.14$, 
we find numerically that the $95.45 \% C.L.$ lower limit is 
reduced from $M_{KK}=4025$ GeV down to $M_{KK}=3055$ GeV if one assumes instead a lower VEV fixed at $v_{SM}=245$ GeV. 
In other words, the increase of the VEV due to included RS corrections has to be compensated by an increase of $M_{KK}$.
In conclusion, the obtained significant increase of the Higgs VEV in RS [see the effect by comparing the black point and star in Fig.(\ref{fig:ST})] 
can lead to a large enhancement of the $M_{KK}$ lower limit from EWPT, for given sets of RS parameters (it never translates into a reduction of the limit).

\section{Higgs boson phenomenology}
\label{Higgs}

\subsection{Higgs boson couplings}
\label{couplings}

In order to discuss quantitatively the Higgs boson phenomenology at past, present and future high--energy colliders, 
let us first derive the Higgs couplings within the RS framework.

\subsubsection{Couplings to EW gauge bosons}
\label{hVVsec}

After EWSB, the neutral gauge bosons ($Z^0$ and $Z'$) and their KK excitations get masses through their couplings to the Higgs boson which has a bidoublet structure under the 
custodial symmetry group ${\rm SU(2)_L \! \times \! SU(2)_R}$. These mass terms can be written, in the 4D Lagrangian including the first three KK excitations, as  
$$
{\cal L}_{\rm mass}^{\rm n} \! = \! \frac{1}{2} ( Z^0_\mu \ \ Z^{(1)}_\mu \ \ Z'_\mu \ \ \dots \ \ Z^{(3)}_\mu \ \
Z'^{(3)}_\mu ) \ {\cal M}^2_0  \
( Z^{0\mu} \ \ Z^{(1)\mu} \ \ Z'^\mu \ \ \dots \ \ Z^{(3)\mu} \ \ Z'^{(3)\mu} )^T 
$$
where the squared mass matrix ${\cal M}^2_0$ is given in Eq.(\ref{NeutMassMat}). Numerically, we indeed include the KK tower up to the third excitations $Z^{(3)}$ 
and $Z'^{(3)}$ (included). This effective truncation of the tower induces an error of less than the percent on the $hVV$ couplings computed in this section. 
Similarly, the charged gauge bosons ($W^\pm$ and $\widetilde W^\pm$) get KK--type and VEV--induced masses ({\it c.f.} Eq.(\ref{ChargMassMat})):
$$
{\cal L}_{\rm mass}^{\rm c} \! = \! (W^+_\mu \ \ W^{+(1)}_\mu \ \ \widetilde W^{+}_\mu \ \ \dots \ \ W^{+(3)}_\mu \ \
\widetilde W^{+(3)}_\mu) \ {\cal M}^2_\pm  \
( W^{-\mu} \ \ W^{-(1)\mu} \ \ \widetilde W^{-\mu} \ \ \dots \ \ W^{-(3)\mu} \ \ \widetilde W^{-(3)\mu} )^T .
$$
The mass matrix ${\cal M}^2_0$ is diagonalized by a $7 \times 7$ unitary matrix $U$,
via the basis transformation  $(Z_{\mu} \ \ Z^1_{\mu} \ \ Z^2_{\mu} \ \ \dots )^T \!=\!\ U \ (Z^0_\mu \ \ Z^{(1)}_\mu \ \ Z'_\mu \ \ \dots )^T$ :   
\begin{equation}
\mathcal{M'}^2_0 \equiv U \ {\cal M}^2_0 \ U^\dagger \ = \ {\rm diag}~(m^2_Z,m^2_{Z^1},\dots,m^2_{Z^6}) ,
\label{MnEIGEN}  
\end{equation}
$m_Z$ having to be associated with the experimental value for the $Z$ boson mass (the
unitary matrix is chosen such that $m_Z\!<\!m_{Z_1}\!<\!m_{Z_2}\!<\!\dots$). 
The $Z_{\mu}$ and $Z^i_{\mu}$ [$i=1,2,\dots$] components are the mass eigenstates.  
The mass matrix ${\cal M}^2_\pm$ is diagonalized by a $7 \times 7$ unitary matrix $V$ through,
\begin{equation}
\mathcal{M'}^2_\pm \equiv V \ {\cal M}^2_\pm \ V^\dagger \ = \ {\rm diag}~(m^2_W,m^2_{W^1},\dots,m^2_{W^6}) .
\label{McEIGEN}  
\end{equation}
The values of $\tilde v$, $g$ and $g'$ are obtained by solving the system formed by Eq.(\ref{GF}) and Eq(\ref{MnEIGEN})-(\ref{McEIGEN}) [first eigenvalue].
The $\tilde v$ values obtained that way do not differ significantly from the ones derived using the method developed in Section \ref{system}.
From the obtained values of $\tilde v$, $g$, $g'$, one can now deduce the rotation matrices $U$ and $V$ needed to compute the couplings.

In the weak basis, the 4D Higgs couplings to neutral gauge bosons are then
$$
{\cal L}_{\rm coupling}^{\rm n} \! = \ \frac{h}{\tilde v} \ ( Z^0_\mu \ \ Z^{(1)}_\mu \ \ Z'_\mu \ \ \dots \ \ Z^{(3)}_\mu \
\ Z'^{(3)}_\mu ) \ {\cal C}_0 \
( Z^{0\mu} \ \ Z^{(1)\mu} \ \ Z'^\mu \ \ \dots \ \ Z^{(3)\mu} \ \ Z'^{(3)\mu} )^T 
$$ 
the matrix ${\cal C}_0$ being given in Eq.(\ref{NeutCouplMat}). For the charged gauge bosons [{\it c.f.} Eq.(\ref{ChargCouplMat})],
$$
{\cal L}_{\rm coupling}^{\rm c} \! = \ 2 \ \frac{h}{\tilde v} \ (W^+_\mu \ \ W^{+(1)}_\mu \ \ \widetilde W^{+}_\mu \ \ \dots \
\ W^{+(3)}_\mu \ \ \widetilde W^{+(3)}_\mu) \ {\cal C}_\pm  \
( W^{-\mu} \ \ W^{-(1)\mu} \ \ \widetilde W^{-\mu} \ \ \dots \ \ W^{-(3)\mu} \ \ \widetilde W^{-(3)\mu} )^T .
$$
Moving to the mass basis, the neutral gauge boson interactions are described by the Lagrangian:
\begin{eqnarray}   
{\cal L'}_{\rm coupling}^{\rm n} \  = \ \frac{h}{\tilde v} \ (Z_{\mu} \ \ Z^1_{\mu} \ \ Z^2_{\mu} \ \ \dots ) \ {\cal C'}_0 \
(Z^{\mu} \ \ Z^{1\mu} \ \ Z^{2\mu} \ \ \dots ) ^T  ,  
\ \mbox{where,} \ 
{\cal C'}_0 \ = \ U \ {\cal C}_0 \ U^\dagger .
\label{ZinterMASS}   
\end{eqnarray}
Similarly,
\begin{eqnarray}   
{\cal L'}_{\rm coupling}^{\rm c} \  = \ 2 \ \frac{h}{\tilde v} \ (W_{\mu} \ \ W^1_{\mu} \ \ W^2_{\mu} \ \ \dots ) \ {\cal
C'}_\pm \
(W^{\mu} \ \ W^{1\mu} \ \ W^{2\mu} \ \ \dots ) ^T  ,  
\ \mbox{with,} \ 
{\cal C'}_\pm \ = \ V \ {\cal C}_\pm \ V^\dagger .
\label{WinterMASS}   
\end{eqnarray}
Therefore, the $hZZ$ and $hWW$ effective dimensionful coupling constants calculated within the RS framework are respectively 
the $(1,1)$--matrix elements: $g^{\rm RS}_{hZZ}=({\cal C'}_0 \vert_{11}) / \tilde v$ and $g^{\rm RS}_{hWW}=2 \ ({\cal C'}_\pm \vert_{11}) / \tilde v$. 
In contrast, within the pure SM case, the Higgs coupling constants are 
$g^{\rm SM}_{hZZ}= ( \{g^2+g'^2\} v^2_{SM} / 4 + \delta^{\rm SM}g_{hZZ}) / v_{SM}$ and 
$g^{\rm SM}_{hWW}= 2 \ ( g^2 v^2_{SM} / 4 + \delta^{\rm SM}g_{hWW}) / v_{SM}$
where $g$,$g'$ are calculated from Eq.(\ref{MZ2})-(\ref{MW2}), of course without the RS corrections, and
the $\delta$'s parts represent the SM loop corrections [given in Eq.(\ref{DSMirrZ})-(\ref{DSMirrW})].

\begin{figure}[!ht]
\begin{center}
\includegraphics[width=10.cm]{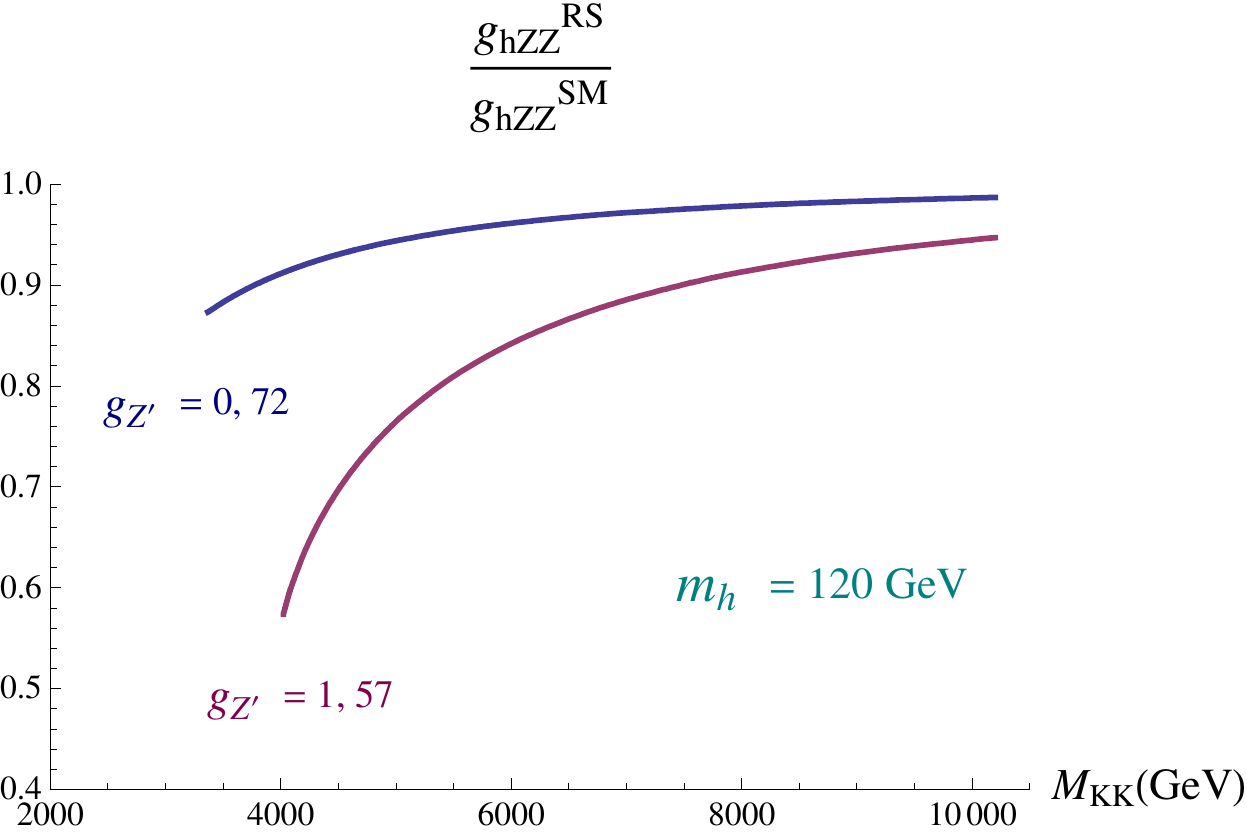} 
\end{center}
\caption{Value of the Higgs--$ZZ$ coupling ratio $g^{\rm RS}_{hZZ}/g^{\rm SM}_{hZZ}$ as a function of $M_{KK}$ [in GeV] for $g_{Z'} = 0.72$ and $1.57$. 
The two minimum values of $M_{KK}$, where the curves stop respectively, are equal to the $M_{KK}$ lower limits from EWPT obtained with $m_h = 120$ GeV 
(points A and B of Table \ref{MainTable}).}
\label{fig:ghVV}
\end{figure}

The obtained values of the ratios $g^{\rm RS}_{hVV}/g^{\rm SM}_{hVV}$ are given in Table \ref{MainTable} for three characteristic points of the parameter space. 
These parameter sets respect the EWPT constraints. In particular, the Higgs boson VEV's $\tilde v$ corresponding to these sets, and involved in $g^{\rm RS}_{hVV}$,
lead to $S_{\rm RS}$ and $T_{\rm RS}$ values within the $95.45 \% C.L.$ contour levels. The obtained $Z$ and $W$ coupling strengths are nearly identical 
because of the approximate custodial symmetry. 
The first conclusion about these numerical results is that the RS corrections of the Higgs boson couplings to EW gauge bosons can be quite strong. Secondly,
the given $\tilde v$ values in the same table show that the role of the Higgs VEV modification in the RS corrections to $g_{hVV}$ is major. 
For the example of point A, the total relative correction on the Higgs coupling to $Z$ is of 
$\delta g_{hZZ}/g_{hZZ}=(g^{\rm RS}_{hZZ}-g^{\rm SM}_{hZZ})/g^{\rm SM}_{hZZ}=- 42.7\%$ and an included correction of $- 23.9\%$ is explained by the VEV: 
assuming {\it only} a variation in the denominator through the VEV (both RS and SM couplings are inversely proportional to the Higgs VEV), the correction is 
indeed $\delta g_{hZZ}/g_{hZZ}=v_{SM}/\tilde v - 1$.
The numbers in Table \ref{MainTable} confirm that $g^{\rm RS}_{hZZ}$ gets more suppression w.r.t. SM as $\tilde v$ increase.  
The remaining part of the whole negative RS correction to $g_{hZZ}$ comes from the direct KK gauge boson mixing effect on the Higgs coupling (numerator part). 
The $g^{\rm RS}_{hZZ}$ suppression is more effective for a larger KK gauge mixing effect on the coupling [i.e. larger $g_{Z'}$ or smaller $M_{KK}$]
as illustrated in the same table and Fig.(\ref{fig:ghVV}). 
In Fig.(\ref{fig:ghVV}), we plot the ratio $g^{\rm RS}_{hZZ}/g^{\rm SM}_{hZZ}$ for RS parameters respecting the EWPT constraints (sets A and B of 
Table \ref{MainTable} with larger $M_{KK}$'s). It shows that RS corrections to the $hVV$ vertex remain important in wide regions of the 
allowed parameter space. 

\subsubsection{Yukawa couplings}
\label{couplYuk}

\noindent
\textbf{Yukawa couplings with KK mixings:} The fermions mass matrices need to be studied as the Yukawa couplings 
induce mixing with the KK excitations. Moreover, the Yukawa terms have to take an invariant form under the custodial symmetry 
as the Higgs field is embedded into a bidoublet of the ${\rm SU(2)_L\! \times\! SU(2)_R}$ gauge symmetry. As a consequence, 
fermions are promoted to higher gauge multiplets, the simplest realization being the right-handed fermion singlets promoted to 
${\rm SU(2)_R}$ isodoublets, and new exotic quarks are expected. Since the ${\rm SU(2)_R}$ symmetry is broken by boundary conditions, 
these exotic quarks have $(-+)$ boundary conditions (BC), so that those have no zero--mode, in contrast with the SM fermions 
which have $(++)$ BC. Therefore, the first KK excitation of those exotic quarks can be relatively low depending on the value 
of the $c_f$ parameter.

Defining $\mathcal{M}_f$ as the mass matrix of a specific chiral fermion $\Psi_{L/R}$ in the interaction basis
(see e.g. Appendix \ref{FermMat}), 
this matrix can be diagonalized by unitary matrices $U_{L/R}$, through the transformation $\Psi'_{L/R}=U_{L/R}\Psi_{L/R}$ :
\begin{eqnarray}
\mathcal{M}'_f \equiv U_L \mathcal{M}_f U^{\dagger}_R = {\rm diag}~(m_{f_{1}} ,m_{f_{2}} ,\dots )
\label{MfDIAGO}
\end{eqnarray}
where $m_{f_{1}}$ corresponds to the measured value of the SM fermion mass 
(the unitary matrices being defined such that $m_{f_{1}} < m_{f_{2}} < . . .$). The $\Psi'_{L/R}$ components are the mass 
eigenstates. Then, the Higgs couplings are given by the interaction Lagrangian
\begin{eqnarray}
{\cal L}_{int} = {h\over \tilde{v}}\bar{\Psi}'_L \mathcal{C}'_f \Psi'_R + \mbox{h.c.}
\label{CfDIAGO}
\end{eqnarray}
where $\mathcal{C}'_f=U_L\mathcal{C}_f U^{\dagger}_R$ with $\mathcal{C}_f$ being the same matrix as $\mathcal{M}_f$ 
but with the KK masses set to zero (see for instance Appendix \ref{FermCoupl}). 
Hence, the 4D effective Higgs coupling to an eigenstate $f_i$ (i.e. a component of $\Psi'$) is given by 
$\lambda_{f_i}^{\rm RS} = (\mathcal{C}'_{f}|_{ii})/\tilde{v}$ in contrast to the usual $m_f/v_{SM}$ Yukawa value within the SM. 
\\ \\ \noindent
\textbf{Light fermion sector:} For light fermions (other than the $b$- and $t$-quarks), the $c_{\rm light}$ parameters controlling each 
fermion multiplet localization along the fifth dimension need to be higher than $0.5$ in order to reproduce the correct set of 
masses. As a consequence, wave function overlaps between light modes, and their KK excitations or exotic KK partners, 
induce negligible mixings (of order $(m_{f}/M_{KK})^2$) over the zero mode contribution. 
Hence, one can safely neglect the KK fermion contributions, inducing a 4D effective Yukawa coupling for light fermions being 
$\lambda_{f_{\rm light}}^{\rm RS} \simeq m_{f_{\rm light}}/\tilde{v}$. 

In the end, the deviation of the Yukawa coupling in the RS scenario with respect to the SM for light fermions 
simplifies to the deviation of the Higgs boson VEV determined in Section \ref{system}. It gives:
\begin{eqnarray}
{{\lambda_f^{\rm RS}}\over{\lambda_f^{\rm SM}}} \simeq {{v_{SM}}\over{\tilde{v}}} 
\quad\mbox{for every light fermions (other than the $b$- and $t$-quarks).}
\label{lightYUK}
\end{eqnarray}
\\ \noindent
\textbf{Third quark generation sector:} What differs greatly between the light fermion sector and the third quark generation sector 
is the $c_{b,t}$ parameters controlling the bottom/top localization. Indeed, in order to reproduce the top and bottom masses, 
one needs $c_{b,t}$ parameters smaller than $0.5$. Having so, in particular,
the wave function overlaps between the zero--modes and their KK excitations over 
the Higgs field on the TeV--brane can become quite large, inducing possibly large KK mixing effects in the Yukawa couplings.

The KK mixing depends on the mass matrix and hence on the quark representations under the gauge symmetry.
In the sector of third generation quarks, the crucial EW constraints come from the precise measurements of the observables 
$A^b_{FB}$ and $R_b$ (ratio of the partial decay width into $b \bar b$ for the $Z^0$ boson). 
To be complete in our study of the Higgs boson phenomenology with regard to all the EWPT constraints, we will follow the previous work
\cite{RSrep} where we have determined quark representations able to address the $A^b_{FB}$ anomaly 
(i.e. improve significantly the global fit of $R_b$ and the $A^b_{FB}$ values at the $Z^0$ pole and outside resonance).  
Doing so, we consider the proposed choice of multiplets under the ${\rm SU(2)_L\! \times\! SU(2)_R\! \times\! U(1)_X}$ 
group for the bottom and top quarks:
\begin{eqnarray}
&& \{Q_{1L}\} \equiv (2,2)_{2/3} = \left ( \begin{array}{cc} q_{(5/3)L}' & t_{1L} \\ t_L' & b_{1L} \end{array} \right ) 
\quad\quad\quad\quad \{t_R^c\} \equiv (1,3)_{2/3} = \left ( \begin{array}{ccc} q^{c\prime}_{(5/3)R} & t^c_R & b^{c\prime}_R \end{array} \right ) 
\nonumber\\
&& \{Q_{2L}\} \equiv (2,3)_{-5/6} = 
\left ( \begin{array}{ccc} t_{2L} & b_L' & q_{(-4/3)L}'' \\ b_{2L} & q_{(-4/3)L}' & q_{(-7/3)L}' \end{array} \right ) 
\quad \{b_R^c\} \equiv (1,2)_{-5/6} = 
\left ( \begin{array}{ccc} b^c_R & q^{c\prime}_{(-4/3)R} \end{array} \right ) 
\label{Model}
\end{eqnarray}
where the left bottom and top SM-like multiplets arise from a mixing mechanism between the two above left-handed multiplets
(parametrized by a mixing angle $\theta$)
\footnote{The precise description for the mentioned mechanism is left to the attention of the reader in our previous paper \cite{RSrep}.}. 
The heavy quarks indicated with one or several primes (and an electric charge) are the mentioned fields with $(-+)$ BC
[the so--called `custodians']. For the rest of the paper, all numerical values are based on this quark model.

In the field basis $\Psi^t_L \equiv (b^{(0)}_L, b^{(1)}_L, b^{c(1)}_L, b''^{(1)}_L, b'^{c(1)}_L)^t$, 
$\Psi^t_R \equiv (b^{c(0)}_R, b^{(1)}_R, b^{c(1)}_R, b''^{(1)}_R, b'^{c(1)}_R)^t$ where we have introduced the charge 
conjugated fields (indicated by the superscript $c$) in order to use only left-handed SM fields, the effective 
4D bottom quark mass matrix and Yukawa coupling matrix induced by EWSB are given in Appendix \ref{FermMat} and \ref{FermCoupl}. 
Similar structure can be found for the top quark mass and 
Yukawa coupling matrices.  
\\ \\ \noindent
\textbf{Parameter space:} 
Our choice of parameters in order to derive the deviation of the Yukawa couplings relies on a few considerations 
we discuss now.

First, numerically we consider only the first two fermionic custodian excitations, which gives rise to heaviest eigenvalues 
of $\mathcal{M}_f$ around $\sim 2 M_{KK}$ in agreement with the NDA estimation of the cut--off scale of the effective field theory
related to the perturbativity of Yukawa couplings: $\Lambda_{IR} \sim 2 M_{KK}$. 

Secondly, in order to derive the correct value of Yukawa couplings in the RS model, we need to implement in Eq.(\ref{MfDIAGO})
the good value of the relevant fermion mass $m_{f_{1}}$ and the exact unitary matrices $U_{L/R}$. In particular,
the whole $b$- and $t$-quark mixings with the first two quark families are treated effectively through the parameters 
$\epsilon^{mixing}_{b,t}$ appearing in front of the first element of the mass matrices $\mathcal{M}_{b,t}$ (see Appendix \ref{FermMat}). 
As the Cabibbo-Kobayashi-Maskawa (CKM) 
quark mixing matrix $V_{CKM}= U^{up}_L U^{down \dagger}_L$ is close to unity, the simplest case corresponds to both 
rotation matrices for up and down fields being also close to unity. We thus estimate $\epsilon^{mixing}_{b,t}$ to be roughly 
of order $\cos \theta_{12} \cos \theta_{13} \sim \cos \theta_{12} \cos \theta_{23} \sim 0.97$ 
where $\theta_{ij}$ are the CKM mixing angles encoding its hierarchical structure. We then use the allowed range 
$\epsilon^{mixing}_{b,t} = [0.95 , 1.05]$ for the numerical analysis. 

A more important effect comes from the mass running and is also implemented into 
$\epsilon_{b,t} = \epsilon_{b,t}^{mixing} \times \epsilon_{b,t}^{running}$.
The parameters $c_{b,t}$ and the dimensionful Yukawa coupling constants $\lambda^{5D}_{b,t}$, entering $\mathcal{M}_{b,t}$, 
are parameters appearing in the 5D Lagrangian and have thus to be considered e.g. at the effective 5D scale. 
Then, one has to consider the running of quark masses from $M_{KK}$ typically down
to the EWSB scale $\Lambda_{EW} \sim m_Z$ where the EWPT and light Higgs phenomenology are studied in the present paper.
The approximate effect of such a running can be estimated from the mass values at the two extreme scales:
\begin{eqnarray}
&&m_b(m_Z) = [2.8 , 3.0] \mbox{GeV} ,\quad m_b(10 \mbox{TeV}) = [2.1 , 2.3] \mbox{GeV} \nonumber\\
&&m_t(m_Z) = [168 , 180] \mbox{GeV} ,\quad m_t(10 \mbox{TeV}) = [140 , 148] \mbox{GeV} \nonumber
\end{eqnarray}
where we have used the SM one loop renormalization group equations \cite{ref22} to run the quark masses given in Ref.~\cite{ref23} 
from the scale $m_Z$ to $10$ TeV. At that point, one can neglect KK loop contributions to the running which would 
correspond to higher order corrections. The running reduces at most the quark masses by about 
$21 \%$-$42 \%$ ($13 \%$-$28 \%$) for the bottom (top) mass and is taken into 
account through the considered allowed range for the whole factors $\epsilon_{b,t}$:
\begin{eqnarray}
\epsilon_b = [1.15 , 1.45] \quad,\quad \epsilon_t = [1.10 , 1.35] . \nonumber
\end{eqnarray}

Anyway, the precise variation of the $\epsilon_{b,t}$ value has no 
important effects on the final Yukawa coupling, as the rotation matrices $U_{L/R}$ in Eq.(\ref{MfDIAGO}) 
are systematically such that the smallest mass eigenvalue is equal to $m_{f_{1}}$, namely the measured fermion mass.
Nevertheless, for completeness, we include these small possible variations of $\epsilon_{b,t}$. 
Those lead to several possible values of the parameters $c_{b,t}$, $\lambda^{5D}_{b,t}$ and
the mixing angle $\theta$ which reproduce the correct $m_b(m_Z)$, $m_t(m_Z)$ and address the $A^b_{FB}$ anomaly 
[the $A^b_{FB}$ solution is fixed by $\mathcal{M}_b$ which determines $b$-$b^{KK}$ mixings]. 
Here we are clearly thinking for a given value of $M_{KK}$ which also enters $\mathcal{M}_{b,t}$ 
through the dependencies of KK fermion masses. 
Once all parameters are fixed, the Yukawa couplings are extracted through the method given in Eq.(\ref{CfDIAGO}).
\\ \\ \noindent
\textbf{Results and discussion:} The obtained values of the ratios $\lambda^{\rm RS}_{b,t}/\lambda^{\rm SM}_{b,t}$ are 
given in Table \ref{MainTable} for three characteristic points of the parameter space respecting the EWPT constraints
in the light fermion and gauge boson sector. For these three fixed values of $M_{KK}$, the parameters $c_{b,t}$, 
$\lambda^{5D}_{b,t}$ and $\theta$ are varied as described above according to $m_{b,t}(m_Z)$ and $A^b_{FB}$.
Those variations give rise to a certain range of 4D Yukawa coupling constants, for which the extremal values
are given in Table \ref{MainTable}.

The first conclusion about these numerical results is that the RS corrections 
of the Higgs couplings to fermions can be quite strong as was the case for the Higgs coupling to the EW gauge bosons 
(up to $-23.8\%$ for the example of point A). However, in contrast with the EW gauge boson couplings,  
comparing here RS corrections to the $b$- and $t$-quarks with light 
fermions, the table shows that the role of the Higgs VEV modification in the RS corrections to $\lambda_{b,t}$ is major 
in regard of KK mixing effects. At the same time, one can see that the KK mixing corrections also tend to decrease 
the coupling of the Higgs boson to a few more percents (up to $-5.5\%$ for the example of point B) which is what one 
can naively expect: for a given fermion mass value, the higher the KK mixing component is, the lower the Yukawa coupling is
(`direct' mass contribution).

\subsubsection{Effective coupling to gluons}

\begin{figure}[!ht]
\begin{center}
\includegraphics[width=6.cm]{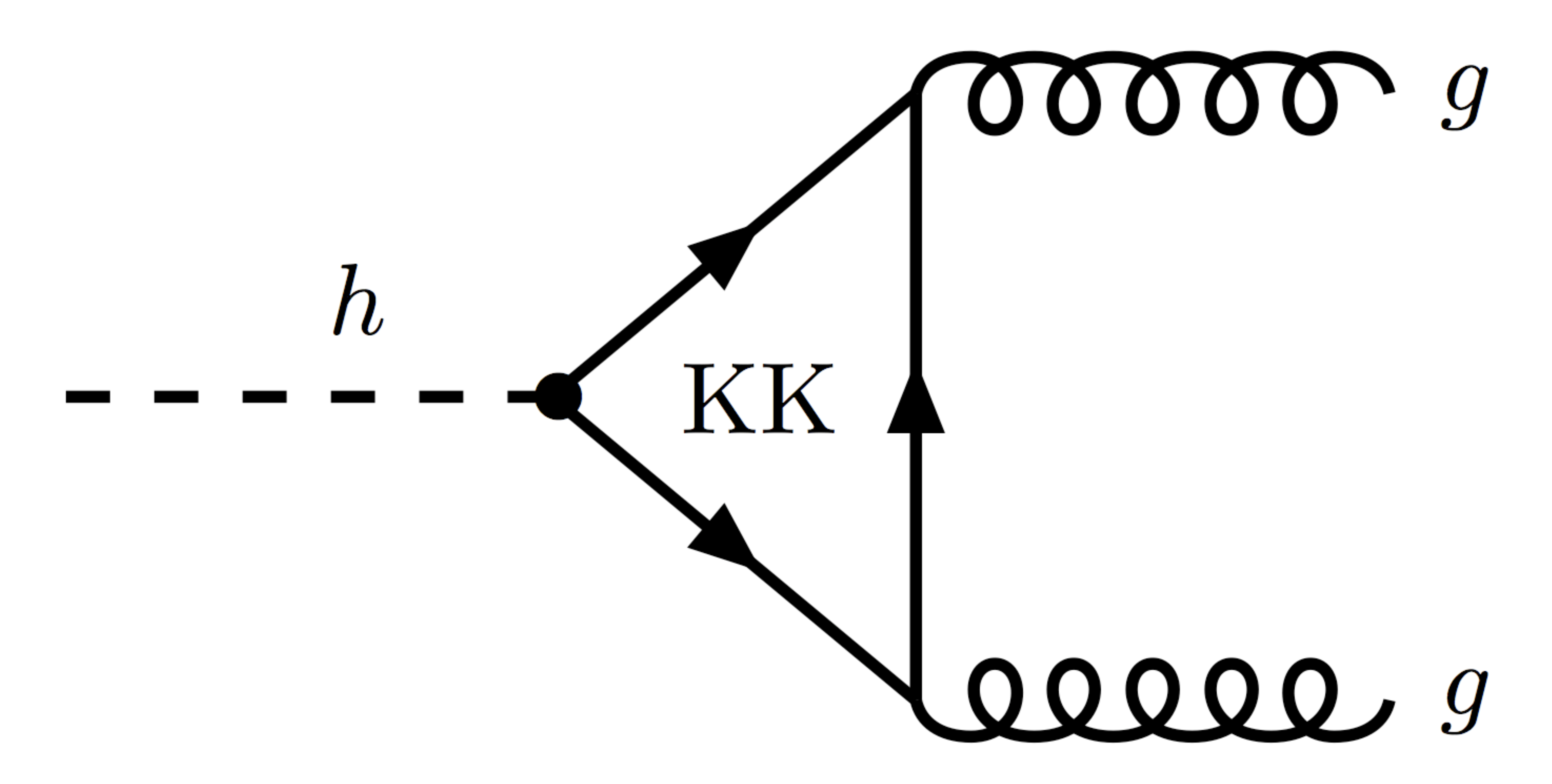} 
\end{center}
\vspace*{-5mm}
\caption{Feynman diagram for the loop-induced Higgs-gluon-gluon vertex.
KK represents the exchanged zero--modes and KK towers of quarks.}
\label{fig:Dhgg}
\end{figure}

\noindent
\textbf{Gluon fusion in the SM:} The dominant production mode of the SM Higgs boson at the LHC 
is the reaction $gg \to h$ called the gluon--gluon fusion mechanism and mediated by triangular loops of SM quarks (noted as $Q$'s here),
as illustrated in Fig.~(\ref{fig:Dhgg}). 
In the SM, mainly heavy quarks, namely the top quark and to a lesser extent the bottom quark, contribute to the amplitude. 
The decreasing Higgs form factor with rising loop mass is counterbalanced by the linear growth 
of the Higgs coupling with the quark mass (Yukawa coupling).

To lowest order, the partonic cross section reads as
\begin{eqnarray}
\sigma_h^{SM} &=& {{\alpha_s^2 m_h^2}\over{576 \pi v_{SM}^2}} \left| \sum_Q A^h_{1/2}(\tau_Q) \right|^2 
\label{SMhgg}
\end{eqnarray}
and it must be multiplied by $\delta(\hat{s}-m_h^2)$ where $\hat{s}$ is the $gg$ invariant energy squared. 
The form factor for spin-${1\over2}$ particles is given by
\cite{HiggsReviewI}:
\begin{eqnarray}
A^h_{1/2}(\tau) = {3\over2}[\tau + (\tau-1)f(\tau)]\tau^{-2} , \quad \mbox{where} \quad f(\tau) = \left\{ \begin{array}{ll} 
\arcsin^2{\sqrt{\tau}} & \tau \leq 1\\ 
-{1\over4}\left[ln \ {{{1+\sqrt{1-\tau^{-1}}}\over{1-\sqrt{1-\tau^{-1}}}}}-i\pi\right]^2 & \tau > 1
\end{array} \right.
\label{FFhalf}
\end{eqnarray}
The form factor $A^h_{1/2}(\tau_Q)$ with $\tau_Q={{m_h^2}/{4m_Q^2}}$ is normalized such that for $m_h \ll m_Q$, 
it reaches unity while it approaches zero in the chiral limit $m_Q\rightarrow0$.
\\ \\ \noindent
\textbf{The $g^{\rm RS}_{hgg}$ coupling:} We now extend this result to the RS scenario. 
First, new loop contributions appear from the various KK excitations of usual SM quarks $q$, 
and the ones coming from possible exotic quarks $q'$ with $(-+)$ or $(+-)$ BC, 
extending the sum in Eq.(\ref{SMhgg}) as a consequence over all possible KK quarks coupled to the Higgs field (see Fig.~(\ref{fig:Dhgg})
on the effective $hgg$ coupling). 
Secondly, the decrease in the Yukawa couplings evaluated in previous section will tend to reduce the $hgg$ 
effective coupling through the $hqq$ vertex appearing in the loop. 
In contrast, the $gqq$ ($gq'q'$) vertex couples universally all (KK) quarks with SM-like strength,
due to the flat profile of the gluon fields along the fifth dimension.

Within this setup the production cross section is simply generalized to:
\begin{eqnarray}
\sigma_h^{\rm RS} = {{\alpha_s^2 m_h^2}\over{576 \pi v_{SM}^2}} 
\left| \sum_{\{q\}} {{\lambda^{\rm RS}_q v_{SM}}\over{m_q}} A^h_{1/2}(\tau_q) 
+ \sum_{\{q'\}} {{\lambda^{\rm RS}_{q'} v_{SM}}\over{m_{q'}}} A^h_{1/2}(\tau_{q'}) 
\right|^2 ,
\end{eqnarray}
since e.g. $\lambda^{\rm RS}_q / \lambda^{\rm SM}_q = \lambda^{\rm RS}_q v_{SM} / m_q$.
The sum over $\{q\}$ includes all SM quarks and their KK partner towers, whereas the sum over $\{q'\}$ includes 
all possible custodian quarks of KK type depending on the model under consideration. $\lambda^{RS}_q$ $(\lambda^{RS}_{q'})$ 
denotes the Yukawa coupling of the corresponding quark $q$ $(q')$ to the Higgs field in the mass eigenbasis. 
$m_q$ $(m_{q'})$ is the mass of the corresponding quark $q$ $(q')$ running in the loop.

It is convenient to consider the ratio $\mathcal{R}_{hgg}=\sigma^{\rm RS}_h/\sigma^{\rm SM}_h$ of the 
$gg \to h$ cross sections in the RS and SM models which can be rewritten,
\begin{eqnarray}
\mathcal{R}_{hgg} = \left( {{v_{SM}}\over \tilde{v}} \right)^2 
\left| {{\sum_{\{q\}} {{\lambda^{^{RS}}_q\tilde{v}}\over{m_q}} A^h_{1/2}(\tau_q) 
+ \sum_{\{q'\}} {{\lambda^{^{RS}}_{q'}\tilde{v}}\over{m_{q'}}} 
A^h_{1/2}(\tau_{q'})}\over{\sum_Q A^h_{1/2}(\tau_Q)}} \right|^2 .
\label{eq:hgg}
\end{eqnarray}
Then the higher order QCD corrections, which are known to be rather large \cite{NLOhgg}, are
essentially the same for all quark species and, thus, drop in this ratio.
From this ratio, one can also deduce the effective $hgg$ loop-coupling deviation from the SM prediction by the straightforward relation:
\begin{eqnarray}
{{g^{\rm RS}_{hgg}}\over{g^{\rm SM}_{hgg}}} \equiv \sqrt{\mathcal{R}_{hgg}} .
\end{eqnarray}

We can use the formulas demonstrated in Appendix \ref{FermionSumRule}, more precisely Eq.(\ref{SUMq}) and Eq.(\ref{SUMqprime}), 
in order to simplify our expression. The ratio $\mathcal{R}_{hgg}^{1/2}$ then simplifies to:
\begin{eqnarray}
{{g^{\rm RS}_{hgg}}\over{g^{\rm SM}_{hgg}}} = {v_{SM}\over \tilde{v}} 
\left| {{\sum_Q \left( 1 + {{\lambda^{^{RS}}_Q\tilde{v}}\over{m_Q}} [ A^h_{1/2}(\tau_Q) - 1 ] \right)}\over{\sum_Q A^h_{1/2}(\tau_Q)}} \right|
\end{eqnarray}
where the sum appearing in the numerator has been reduced to the SM quarks only. 
It is a remarkable feature that the contributions coming from all KK partners simplify 
to properties over the corresponding zero--mode. In the case of exotic quarks, the contribution 
vanishes due to the absence of zero--mode, even if these new quarks do couple to the Higgs field.

For light quarks, $A^h_{1/2}(\tau_Q) \rightarrow 0$ quickly and the ratio 
$\lambda^{\rm RS}_Q / m_Q \rightarrow 1 / \tilde{v}$ as can be deduced from Eq.(\ref{lightYUK}). 
Hence, their contributions to $\mathcal{R}_{hgg}^{1/2}$ tend to vanish as was already the case in the SM.

For heavy fermions ($b$- and $t$-quarks), overlaps between the zero--mode profile and KK wave functions can be quite large, 
as we have seen in the previous section, so that $\lambda^{\rm RS}_Q / m_Q \neq 1 / \tilde{v}$. 
Besides, considering light Higgs masses of $120$ GeV and $150$ GeV, one check numerically that $|A^h_{1/2}(\tau_b)| \simeq 0.1$ 
which is negligible, but only as a first approximation, relatively to $|A^h_{1/2}(\tau_t)| \simeq 1.05$.

In the end, the deviation in the coupling simply reads:
\begin{eqnarray}
{{g^{\rm RS}_{hgg}}\over{g^{\rm SM}_{hgg}}} \simeq {{v_{SM}}\over \tilde{v}} 
\left| {{ 2 - x_b (1 - A_b) - x_t (1 - A_t)}\over{A_b + A_t}} \right| 
\label{CompleteForm}
\end{eqnarray}
where $A_Q \equiv A^h_{1/2}(\tau_Q)$ and $x_Q \equiv \tilde{v} \lambda^{\rm RS}_Q/m_Q$. 
Note that $x_Q\in[0,1]$. The limit $x_Q\rightarrow 1$ corresponds to the pure Higgs mass case when there is no mixing 
between the fermion zero--mode and its KK partners. $x_Q$ tends to decrease as this mixing gets stronger. 
For a light Higgs mass, so that ${{m_h^2}/{4m_t^2}} \ll 1$, in the limit where $|A^h_{1/2}(\tau_t)| \rightarrow 1$ 
and neglecting the bottom quark contribution: $|A^h_{1/2}(\tau_b)| \rightarrow 0$, our relation gets the following
really simple structure,
\begin{eqnarray}
{{g^{\rm RS}_{hgg}}\over{g^{\rm SM}_{hgg}}} \approx {{v_{SM}}\over \tilde{v}} (2-x_b) .
\label{SimpleForm}
\end{eqnarray}
We present this final approximated relation to help the reader in getting an intuition on the main behavior of 
RS corrections to the Higgs coupling, but numerically, the complete formula (\ref{CompleteForm}) is used.
As discussed in part \ref{system}, $v_{SM} < \tilde{v}$, systematically, which tends to reduce the 
production cross section of the Higgs field at LHC. This consequence is simply due to the fact that 
for a higher value of the Higgs VEV, one needs a smaller value of the Yukawa coupling to reproduce a given fermion mass. 
The second term in Eq.(\ref{SimpleForm}) is less trivial and encodes the whole $b$- and $t$-quark KK towers contribution 
minus the pure $b$-quark contribution. This contribution increases the deviation of the $g^{\rm RS}_{hgg}$ w.r.t. 
$g^{\rm SM}_{hgg}$ as the mixing between the bottom zero--mode with its KK excitations grows. On the contrary,
when $x_b$ decreases, the contribution from the bottom KK quarks becomes more and more crucial to the $hgg$ coupling. 
\\ \\ \noindent
\textbf{Results and discussion:} We have derived the values of the ratio $g^{\rm RS}_{hgg} / g^{\rm SM}_{hgg}$ 
from Eq.(\ref{CompleteForm}) and given them in Table \ref{MainTable} for the same characteristic points of 
parameter space respecting all EWPT constraints (including the $A^b_{FB}$ solution). 
The allowed variations of fundamental parameters give rise to some intervals of values for the Yukawa couplings and 
masses of the KK fermion towers. Based on those intervals, we have determined the maximum and minimum amplitudes
for the loop-induced observable $g^{\rm RS}_{hgg} / g^{\rm SM}_{hgg}$.

Once more, we remark that the RS corrections to the effective Higgs boson coupling to two gluons are possibly 
quite strong (up to $-22.4\%$ for point A). Furthermore, comparing these RS corrections (obeying Eq.(\ref{SimpleForm}) 
in a good approximation) with the case of light fermion coupling to the Higgs boson ({\it c.f.} Eq.(\ref{lightYUK})), 
one concludes again on the major role of the Higgs VEV modification. 
Simultaneously, one can see that the KK mixing corrections combined with the new contributions from exchanges of KK states in the loop 
[synthesized in the $(2-x_b)$ factor effect on Eq.(\ref{SimpleForm})] 
tend to counter the effect of the Higgs VEV deviation, but at a smaller rate.

\subsubsection{Effective coupling to photons}

\begin{figure}[!ht]
\begin{center}
\includegraphics[width=14.cm]{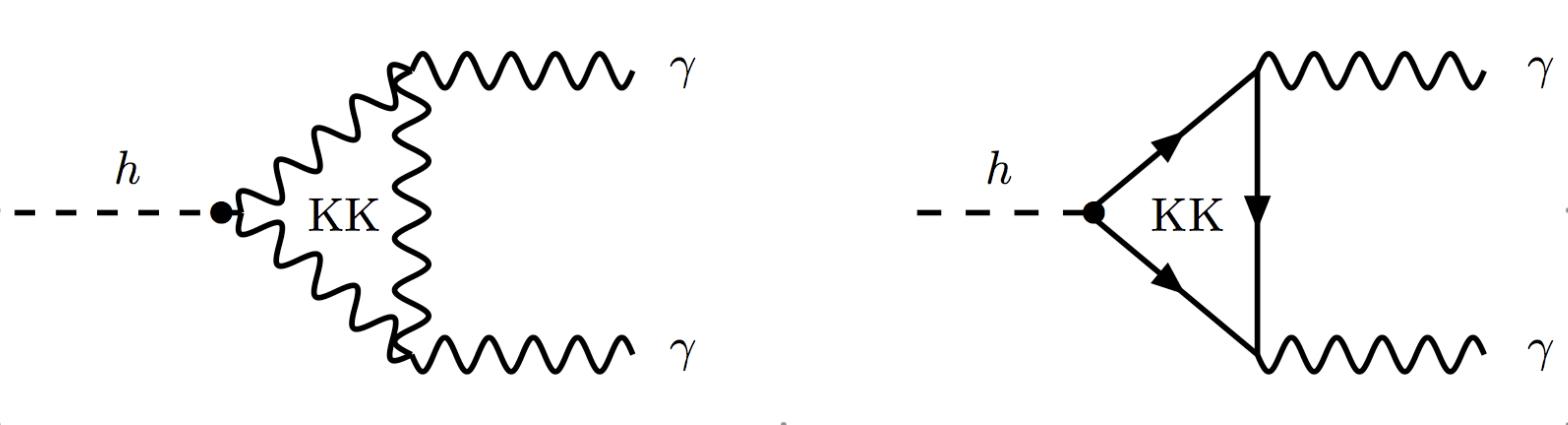} 
\end{center}
\vspace*{-5mm}
\caption{Feynman diagrams for the loop-induced Higgs-photon-photon vertex. 
KK stands for the exchanged zero--modes and KK towers of $W$ gauge boson [left] or fermions [right].}
\label{fig:Dhgaga}
\end{figure}

\noindent
\textbf{$\gamma\gamma$ channel in the SM:} For low Higgs masses, the dominant decay mode $h \to b\bar{b}$ 
is swamped by a large QCD background and the Higgs boson can be searched for through more promising loop-induced decays. 
The decay channel into two photons is the most important one and is mediated by triangular loops of charged fermions as well 
as massive vector bosons: see Fig.~(\ref{fig:Dhgaga}).

The decay width of the Higgs in two photons reads \cite{HiggsReviewI}
\begin{eqnarray}
\Gamma_{h \to \gamma\gamma}^{SM} = {{\alpha^2 m_h^3}\over{256 \pi^3 v_{SM}^2}} 
\left| A^h_{1}(\tau_W) + {4\over3}\sum_f N_c Q_f^2 A^h_{1/2}(\tau_f) \right|^2
\end{eqnarray}
where $N_c$ is the number of color states (3 for quarks, 1 for leptons) and $Q_f$ is the electric charge of 
the fermion in the loop. The form factor for spin-$1/2$ particles, $A^h_{1/2}$, is the one from Eq.(\ref{FFhalf}), 
and the form factor for spin-${1}$ particles is given by:
\begin{eqnarray}
A^h_{1}(\tau) = -[2\tau^2 + 3\tau + 3(2\tau-1)f(\tau)]\tau^{-2} .
\end{eqnarray}
The form factor $A^h_{1}(\tau_V)$ with $\tau_V={{m_h^2}/{4m_V^2}}$ is defined such that for large masses of the boson in the loop, 
$m_h \ll m_V$, it reaches $A^h_{1}(\tau_V \rightarrow 0) = -7$.
\\ \\ \noindent
\textbf{The $g^{\rm RS}_{h\gamma\gamma}$ coupling:}
Here we extend this result to RS case. This extension is similar to the one of the gluon fusion mechanism from previous section. 
New loop contributions appear from the various exchanged KK excitations of usual SM fermions, 
the exchanged custodians, but also the exchanged KK EW gauge bosons (see in Fig.~(\ref{fig:Dhgaga}) 
the diagram for the induced $h\gamma\gamma$ vertex). 
We have to take into account as well the deviations in the Yukawa couplings and in the Higgs coupling to the $W$ boson. 
Finally, the $ff\gamma$, like the $WW\gamma$, vertex couples with an SM-like universal strength all fermion excitations, 
respectively all $W$ bosons of the tower, due to the flat wave function of the electromagnetic field. 
The decay width becomes thus,
\begin{eqnarray}
\Gamma_{h \to \gamma\gamma}^{\rm RS} = {{\alpha^2 m_h^3}\over{256 \pi^3}}{1\over{v_{SM}^2}} 
\left| \sum_{n} {{g^{\rm RS}_{W^n} v_{SM}}\over{2m_{W^n}^2}} A^h_{1}(\tau_{W^n}) + 
{4\over3}\sum_{\{f\}} {{\lambda^{\rm RS}_f v_{SM}}\over{m_f}} N_c Q_f^2 A^h_{1/2}(\tau_f) \right|^2
\end{eqnarray}
as $g^{\rm RS}_{W^n} / g^{\rm SM}_{hWW} = g^{\rm RS}_{W^n} v_{SM} / 2 m_{W^n}^2$ with 
$g^{\rm RS}_{W^n} = g^{\rm RS}_{hW^nW^n} = 2 ({\cal C'}_\pm \vert_{nn}) / \tilde v$, following the notations of Section \ref{hVVsec}.
We recall that $m_{W^n}$ denotes the physical $W$ state eigenmasses and we mention that the KK sum over $n$ includes the zero--mode,
namely the observed $W$ boson.

We still consider the ratio $\mathcal{R}_{h\gamma\gamma} = \Gamma_{h\to\gamma\gamma}^{\rm RS} / \Gamma_{h \to\gamma\gamma}^{\rm SM}$:
\begin{eqnarray}
\mathcal{R}_{h\gamma\gamma} = \left( {v_{_{SM}}\over \tilde{v}} \right)^2 
\left| {3\over 4}{\sum_{n} {{g^{\rm RS}_{W^n}\tilde v}\over{2m_{W^n}^2}} A^h_{1}(\tau_{W^n}) 
+ \sum_{\{f\}} {{\lambda^{\rm RS}_f \tilde{v}}\over{m_f}} N_c Q_f^2 A^h_{1/2}(\tau_f)}\over{{3\over 4}A^h_{1}(\tau_W) 
+ \sum_f N_c Q_f^2 A^h_{1/2}(\tau_f)} \right|^2
\label{eq:hgaga}
\end{eqnarray}
from which one can deduce the $h\gamma\gamma$ coupling deviation,
\begin{eqnarray}
{{g^{\rm RS}_{h\gamma\gamma}}\over{g^{\rm SM}_{h\gamma\gamma}}} \equiv \sqrt{\mathcal{R}_{h\gamma\gamma}}
\end{eqnarray}

Combining now the formulas derived in Appendix \ref{FermionSumRule} and Appendix \ref{BosonSumRule}, 
one easily find in a similar way for the bosonic part:
\begin{eqnarray}
{{g^{\rm RS}_{h\gamma\gamma}}\over{g^{\rm SM}_{h\gamma\gamma}}} \simeq {v_{SM}\over \tilde{v}} 
\left| {{43\over 4} - {9\over4 } x_W (A_W + 7) + x_b (1 - A_b) + 4 x_t (1 - A_t)}\over{{9\over4}A_W + A_b + 4 A_t} \right|
\end{eqnarray}
with $x_W \equiv \tilde{v} g^{\rm RS}_{hWW}/2m_W^2$.
Similarly to the case of gluon fusion Higgs production,
contributions from light quarks and charged leptons vanish, whereas the KK tower contributions for the $W$ boson, quarks 
and charged leptons can be rewritten leaving only in the formula the explicit 
dependence on the $W$ boson, bottom and top quark zero--modes.
\\ \\ \noindent
\textbf{Results and discussion:} The extremal values of ${g^{\rm RS}_{h\gamma\gamma}}/{g^{\rm SM}_{h\gamma\gamma}}$ are also shown
in Table \ref{MainTable}. The $x_{W,b,t}$ values were obtained as discussed in previous sections. Due to the additional $W$
mode effects, the deviation of this effective coupling can reach higher RS corrections (up to $-25.1\%$ e.g. with point A) than $g_{hgg}$. 
The principal RS deviation comes again from the modified value of the Higgs field VEV which is a general result of the present work.

\subsection{The Higgs boson at colliders}
\label{pheno}

\subsubsection{LEP2}

The four LEP collaborations (ALEPH, DELPHI, L3 and OPAL) have used the collected $e^+e^-$ collision data at center-of-mass energies between $91$ GeV and $210$ GeV
to search for the Higgs boson through the Higgs--strahlung production mechanism $e^+e^- \to Zh$ \cite{LEP114}. By exploring the main $b \bar b$ and $\tau^+ \tau^-$ channels,
an upper limit at $95 \% C.L.$ has been put on the products of normalized generic squared couplings and branching ratios: $(g_{hZZ}/g^{\rm SM}_{hZZ})^2 \times B(h \to b \bar b)$ and 
$(g_{hZZ}/g^{\rm SM}_{hZZ})^2 \times B(h \to \tau^+ \tau^-)$ for Higgs masses between $10$ GeV and $120$ GeV. Our point C at $m_h=150$ GeV is thus not excluded by this 2D
constraint. It is also the case for points A and B at $m_h=120$ GeV which correspond to RS values of the products of normalized squared couplings and branching ratios smaller than unity 
(in both channels) since $(g^{\rm RS}_{hZZ}/g^{\rm SM}_{hZZ})^2 < 1$ as we have shown.

Furthermore, assuming even a lighter Higgs boson of say $\sim 99$ GeV [see next paragraph for motivations of this precise choice] 
is still allowed within the RS scenario. First, such a Higgs mass is allowed at 
$95 \% C.L.$ by EWPT together with e.g. $g_{Z'}=1.57$, $M_{KK}=3985$ GeV and $\tilde M/k = 0.09$ (in this section, we will discuss this new set of parameters denoted as the point D).
Secondly, this Higgs mass satisfies the lower bound coming from the considerations on the vacuum stability \cite{HiggsReviewI}. Finally, the above LEP2 upper constraints from both channels
at $m_h \simeq 99$ GeV are also respected for the point D, in contrast with the SM, as then $(g^{\rm RS}_{hZZ}/g^{\rm SM}_{hZZ})^2=0.30$ is much smaller than one so that the Higgs production rate is 
sufficiently reduced. Interestingly, such a light Higgs boson would be difficult to discover at LHC, and the high precision performances of $e^+e^-$ linear colliders would probably be needed.

The combined data of the four LEP collaborations result in an excess of events at $2.3$ standard deviations in the $Z + b's$ channel \cite{LEP114}. 
We claim that it can be nicely fitted by a light Higgs boson 
of $\sim 99$ GeV produced within the RS scenario and decaying into $b \bar b$. Indeed, for the point D, we obtain $\tilde v =337$ GeV and the value 
$(g^{\rm RS}_{hZZ}/g^{\rm SM}_{hZZ})^2 \times B^{\rm RS}(h \to b \bar b)
\simeq 0.231$ which is (smaller but) close to the observed limit of $\sim 0.236$. This correct value of the product is reached within RS thanks to the suppression of $g_{hZZ}$ and to the precise amount of $B(h \to b \bar b)$. The amount of $B(h \to b \bar b)$ is fixed by the 4D effective $b$--quark Yukawa couplings, and hence by its wave functions along the fifth dimension, $M_{KK}$ as well as the 
5D Yukawa coupling constants. In turn e.g. the $b$--quark Yukawa couplings determine the $b$ mixing with its KK excitations, and thus the $A^b_{FB}$ prediction. The point D is chosen such that it 
corresponds to certain $b$--quark wave functions giving rise to the 4D Yukawa coupling ($\lambda^{\rm RS}_b/\lambda^{\rm SM}_b=0.70$) which simultaneously gives the wanted 
$B(h \to b \bar b)$ and $A^b_{FB}$ [+ $R_b$, $m_{b,t}$] values.
We thus underline the non--trivial result that the RS model can both address the LEP1-2 anomaly on $A^b_{FB}$ and explain the LEP2 excess of $Z + b's$ events, 
with common sets of fundamental parameters in the $b$--quark sector in particular.

\subsubsection{Tevatron Run II}

Based on collected data at $\sqrt{s}=1.96$ TeV with ${\cal L}=0.9-4.2$ fb$^{-1}$, 
the combined CDF and D0 analyzes allow to put upper limits at $95 \% C.L.$ on the rates of Higgs production (and decays) for $100$~GeV~$< m_h < 200$~GeV \cite{RunIIHiggs}.
Both experiments have performed dedicated searches in different channels. At high mass, like for our point at $m_h=150$ GeV, all the sensitivity comes from the channel
$gg \to h \to WW$. Within the RS scenario (point C), this Higgs rate has a reduction, relatively to the SM, in the interval:
\begin{eqnarray}
\frac{ \delta [\sigma_h B(h\to WW)] }{ \sigma_h B(h\to WW) } = \frac{ \sigma^{\rm RS}_h }{ \sigma^{\rm SM}_h }\frac{ B^{\rm RS}(h\to WW) }{ B^{\rm SM}(h\to WW) } - 1
= [-44.8,-39.0] \ \%  
\label{DSBsSB}   
\end{eqnarray}
and $B^{\rm RS}(h\to WW) / B^{\rm SM}(h\to WW)$ is found to be in the range $[0.85,0.88]$ (including pure SM radiative corrections only).
This is obtained for the ranges of values for the Higgs couplings and for the ratio $\sigma^{\rm RS}_h / \sigma^{\rm SM}_h = (g^{\rm RS}_{hgg}/g^{\rm SM}_{hgg})^2$ taken from Table \ref{MainTable}. 
The EWPT constraints are thus satisfied. In conclusion, for the point C, the Higgs rate is below the rate for the SM case which is itself well under the exclusion limit 
\cite{RunIIHiggs} so that the considered Higgs mass of $150$ GeV is clearly allowed by the Tevatron constraints.

Similarly, the RS points A and B are clearly permitted by these Tevatron limits. For $m_h=120$ GeV, the dominant Higgs discovery channel is
$q \bar q \to Wh \to l \nu b \bar b$. For example with the parameter set A, 
the reduction of this Higgs rate is $\delta [\sigma_{Wh} B(h\to b \bar b)] / \sigma_{Wh} B(h\to b \bar b) = [-66.4,-65.1] \% $ 
as $\sigma^{\rm RS}_{Wh} / \sigma^{\rm SM}_{Wh} = (g^{\rm RS}_{hWW}/g^{\rm SM}_{hWW})^2 = 0.33$  
and the $B^{\rm RS}(h\to b \bar b) / B^{\rm SM}(h\to b \bar b)$ ratio is inside $[1.01,1.05]$ ({\it c.f.} Table \ref{MainTable}).
The reductions of the other channels w.r.t. SM are $\delta [\sigma_{Zh} B(h\to b \bar b)] / \sigma_{Zh} B(h\to b \bar b) = [-66.6,-65.3] \%$,
$\delta [\sigma_h B(h\to WW)] / \sigma_h B(h\to WW) = [-62.7,-56.5] \%$ and $\delta [\sigma_h B(h\to \tau \bar \tau)] / \sigma_h B(h\to \tau \bar \tau) = [-34.9,-24.2] \%$. 
Therefore, once more, for point A, all the Higgs rates are below the rates in the SM being themselves under the exclusion limit at $95 \% C.L.$ 
\cite{RunIIHiggs} so that the other considered Higgs mass of $120$ GeV is also allowed by the recent Tevatron data.

In the pure SM, these Tevatron results exclude the range $160$~GeV~$\leq m_h \leq 170$~GeV at the $95 \% C.L.$ \cite{RunIIHiggs}.
To reconsider this mass exclusion within the RS case, one should compute the EWPT limits at the various experimental points in this Higgs mass interval 
and then calculate precisely the corrections to Higgs rates. The Higgs rates within the RS scenario depend on the chosen parameter values.
In particular for the minimum $M_{KK}$ values allowed by EWPT, the Higgs rates in RS are expected to be decreased by more than $\sim 30\%$ w.r.t. SM 
at $m_h = 165$~GeV [given the numerical result in Eq.(\ref{DSBsSB}) for the close mass $m_h = 150$~GeV] 
so that these rates reach regions allowed by Tevatron constraints in the range $160$~GeV~$\leq m_h \leq 170$~GeV, making such Higgs masses realistic again.

\subsubsection{LHC}

\noindent {\bf Higgs boson search:}
The large deviations to the gluon--gluon fusion mechanism (producing the Higgs boson) and to the couplings (fixing its branching ratios) can affect the Higgs search at the Large Hadron Collider (LHC). 
For instance with a SM Higgs mass of $150$ GeV, the significance predicted by the ATLAS collaboration is at $\sim 12.5$ with a center-of-mass energy of $14$ TeV 
and an integrated luminosity of ${\cal L}=10$ fb$^{-1}$ \cite{ATLASsig}. The significance is defined typically as 
${\cal S} = \sigma^{\rm SM}_h B^{\rm SM}_h {\cal L} \epsilon_{sig} / \sqrt{ \sigma_{back} {\cal L} \epsilon_{back} }$ 
in the narrow--width approximation 
[still valid when incorporating the RS corrections in the low/intermediate Higgs mass range] where $\epsilon_{sig}$ 
($\epsilon_{back}$) includes the experimental efficiency
for the signal (background), $\sigma^{SM}_h$ is the cross section 
for the Higgs production process within the SM, $B^{SM}_h$ its branching ratios and $\sigma_{back}$ 
the background cross section. The main Higgs production process
is the gluon--gluon fusion mechanism, its considered decay channel 
is $h \to VV$ and the dominant background is the EW gauge boson production. Moving to the RS case, 
the significance would vary according to 
\begin{eqnarray}
\frac{ \delta {\cal S} }{ {\cal S} } \simeq \frac{ \delta [ \sigma_h B_h ] }{ \sigma_h B_h } = \frac{ \sigma^{\rm RS}_h }{ \sigma^{\rm SM}_h }\frac{ B^{\rm RS}_h }{ B^{\rm SM}_h } - 1.  
\label{Sig}   
\end{eqnarray}
The background is assumed to be identical since the deviations of gauge boson couplings have severe constraints from EWPT. 
We do not include in our discussion the dependence of experimental efficiencies on rates and channels.
For instance with the point C of Table \ref{MainTable}, Eq.(\ref{Sig}) gives $\delta {\cal S} / {\cal S}$ between $\sim - 44 \%$ and $\sim - 39 \%$ since $B^{\rm RS}(h\to VV) / B^{\rm SM}(h\to VV)$
lies in the interval $[0.85,0.87]$. This result is based on range of values for the various Higgs couplings and $\sigma^{\rm RS}_h / \sigma^{\rm SM}_h = (g^{\rm RS}_{hgg}/g^{\rm SM}_{hgg})^2$ 
taken in Table \ref{MainTable} so that the EWPT constraints are well respected. We end up with a significance between
${\cal S}_{\rm RS} = {\cal S} (1 + \delta {\cal S} / {\cal S}) \simeq 12.5 (1 - 0.44) = 7.0$ and ${\cal S}_{\rm RS} \simeq 12.5 (1 - 0.39) = 7.6$
so that the Higgs discovery at $5\sigma$ remains possible within RS, even if the significance is greatly reduced.

Initially, the LHC is expected to run at a lower center-of-mass energy of $\sqrt{s}=10$ TeV at which production rates are reduced by about a factor of two (from
those at $\sqrt{s}=14$ TeV) \cite{LHC2FC}. Assuming such an energy, 
$\delta {\cal S} / {\cal S} \simeq (\sigma^{\rm RS}_h B^{\rm RS}_h) / ( \sqrt{2} \sigma^{\rm SM}_h B^{\rm SM}_h ) - 1$ 
and the significance would further decrease down to ${\cal S}_{\rm RS} \simeq 4.8 - 5.3$ at ${\cal L}=10$ fb$^{-1}$ 
rendering a Higgs boson discovery even more challenging
\footnote{By the time the ATLAS and CMS collaborations will have studied $10$ fb$^{-1}$ of data, the Tevatron should have been completed. In case Tevatron runs in 2011, a total collect of
$10$ fb$^{-1}$ analyzed data per experiment [CDF and D0] is also expected. The latest projections by Tevatron groups conclude from such performances that a $3\sigma$
sensitivity could be achieved for a SM--like Higgs boson in the range $150$~GeV~$\leq m_h \leq 180$~GeV \cite{LHC2FC}. In contrast,
the present results for the considered RS parameter set show that with $10$ fb$^{-1}$ the Higgs boson signature observation at center-of-mass energies 
much lower than $10$ TeV -- like at Tevatron -- are compromised.}.
In the SM, much lower luminosities are required at $10$ TeV for detecting a Higgs boson of $150$ GeV 
(see preliminary studies from both CMS and ATLAS \cite{1OTeVHiggs}).

The present results yield indicative estimates of the LHC sensitivity 
for a chosen point of the RS parameter space. Clearly, for a larger $M_{KK}$ e.g. the
suppression of $\sigma_h$ would be soften making the Higgs boson search easier.
\\ \\
\noindent {\bf RS signature search:}
From the argumentation developed in Appendix \ref{Consider}, 
we have obtained the condition for having a $68\%$ probability to observe at least a $1\sigma$ deviation 
(relatively to the experimental uncertainty), due to RS effects, between the SM prediction for a certain quantity 
${\cal Q}_{\rm SM}$ and its measured central value. Applying this condition (Eq.(\ref{final})) to the product 
of Higgs production and decay rates $\sigma_h B_h$, we get, 
\begin{eqnarray}
\frac{\sigma_h B_h \vert_{\rm RS}}{\sigma_h B_h \vert_{\rm SM}} < \frac{1-\delta [\sigma_h B_h]/\sigma_h B_h}{1+\delta [\sigma_h B_h]/\sigma_h B_h} ,
\label{finalapplied}   
\end{eqnarray}
where for $\delta(\sigma_h B_h)/\sigma_h B_h$ we take the relative experimental accuracy according to the prospects at LHC in the measurement of rates for specific individual channels 
(with $\int {\cal L} dt =30$ fb$^{-1}$ as could be collected after several years of run) \cite{Duhrssen}. 
For example, considering the point C of parameter space as above, we find that this condition is fulfilled for the five channels 
[combining all theoretical predictions and experimental sensitivities]:
$$
\frac{\sigma(qq \to hqq) \vert_{\rm RS}}{\sigma(qq \to hqq) \vert_{\rm SM}} \ \frac{B(h\to WW) \vert_{\rm RS}}{B(h\to WW) \vert_{\rm SM}} 
= [0.30, 0.32]
< \frac{1-0.13}{1+0.13} = 0.76  
$$
$$
\frac{\sigma(qq \to hqq) \vert_{\rm RS}}{\sigma(qq \to hqq) \vert_{\rm SM}} \ \frac{B(h\to ZZ) \vert_{\rm RS}}{B(h\to ZZ) \vert_{\rm SM}} 
= [0.30, 0.31]
< \frac{1-0.49}{1+0.49} = 0.34  
$$
$$
\frac{\sigma(gg \to h) \vert_{\rm RS}}{\sigma(gg \to h) \vert_{\rm SM}} \ \frac{B(h\to WW) \vert_{\rm RS}}{B(h\to WW) \vert_{\rm SM}} 
= [0.55, 0.60] 
< \frac{1-0.12}{1+0.12} = 0.78  
$$
$$
\frac{\sigma(gg \to h) \vert_{\rm RS}}{\sigma(gg \to h) \vert_{\rm SM}} \ \frac{B(h\to ZZ) \vert_{\rm RS}}{B(h\to ZZ) \vert_{\rm SM}} 
= [0.54, 0.60]
< \frac{1-0.24}{1+0.24} = 0.61  
$$
\begin{eqnarray}
\frac{\sigma(qq \to Wh) \vert_{\rm RS}}{\sigma(qq \to Wh) \vert_{\rm SM}} \ \frac{B(h\to WW) \vert_{\rm RS}}{B(h\to WW) \vert_{\rm SM}} 
= [0.31, 0.32]
< \frac{1-0.42}{1+0.42} = 0.40  
\label{finalnumeric}   
\end{eqnarray}
$\sigma(qq \to hqq)$ being the cross section for the Weak Boson Fusion (WBF) mechanism.
Taking the ratio of squared amplitudes allow to include implicitly the SM loop corrections and the calculated branching ratios also include those corrections.
The cross sections for the WBF mechanism, the $Wh$ production and the gluon--gluon fusion process are all significantly reduced w.r.t. 
SM due to the large decrease of the effective couplings $g_{hVV}$ and $g_{hgg}$ within RS (studied in Table \ref{MainTable}).
The branching ratio for the decay channel $h\to VV$ is also reduced in RS since $g_{hVV}$ is more reduced than the bottom Yukawa coupling $\lambda_b$ (as discussed above)
and the two channels $h\to VV^*$,$h\to b \bar b$ dominate for $m_h = 150$ GeV. By the way, the variations of the values given in Eq.(\ref{finalnumeric}) are due in particular to the variation of 
$\lambda^{\rm RS}_b$ given in Table \ref{MainTable}. The significant reductions of the cross sections and branching ratios appearing in Eq.(\ref{finalnumeric}) 
together with the expected LHC performances on $\delta [\sigma_h B_h]/\sigma_h B_h$ allow to satisfy the condition (\ref{finalapplied}), or in other words make the RS corrections visible at LHC.
For instance with the first channel, the condition for having a $95\%$ probability to observe at least a $2\sigma$ deviation, due to RS effects, is even fulfilled:
$$
\frac{\sigma(qq \to hqq) \vert_{\rm RS}}{\sigma(qq \to hqq) \vert_{\rm SM}} \ \frac{B(h\to WW) \vert_{\rm RS}}{B(h\to WW) \vert_{\rm SM}} 
= [0.30, 0.32] 
< \frac{1-2 \times 0.13}{1+2 \times 0.13} = 0.58  
$$
In conclusion, the possibly large RS corrections to the Higgs couplings induce deviations w.r.t. 
SM of the Higgs production and decay rates which could be detected at LHC.
It means that the existence of extra dimensions, and more particularly of the RS model, 
could be probed at the LHC via investigations on the Higgs sector only.

The above conclusion is true for the effective Higgs couplings to gluons and to $W$, $Z$ bosons. In contrast, although the RS corrections to the effective $h\gamma\gamma$ coupling can also be large, 
those cannot be individually tested at LHC for $\int {\cal L} dt =30$ fb$^{-1}$. This is due to the smallness of the branching $B(h\to \gamma\gamma)$ which degrades the
experimental sensitivity on the $\gamma\gamma$ channel. As a matter of fact, let us consider the mass $m_h = 120$ GeV at which the promising CMS sensitivity to this channel is optimum
\cite{CMSsig}. For the point A of parameter space, where the deviation to $g_{h\gamma\gamma}$ coming from KK effects is maximal, one gets, 
\begin{eqnarray}
\frac{g^{\rm RS}_{h\gamma\gamma}}{g^{\rm SM}_{h\gamma\gamma}} = [0.74, 0.75] > \frac{1-\delta g_{h\gamma\gamma}/g_{h\gamma\gamma}}{1+\delta g_{h\gamma\gamma}/g_{h\gamma\gamma}} = 0.52,
\label{finalgamma}   
\end{eqnarray}
where $\delta g_{h\gamma\gamma}/g_{h\gamma\gamma} = 0.31$ is the relative error bar that can be obtained on the Higgs coupling by combining several measurements \cite{DirkZerwas}.
It means that the condition (Eq.(\ref{final})) on the observability of a $1\sigma$ deviation between $g^{\rm SM}_{h\gamma\gamma}$ and its measured central value
is not fulfilled. This constitutes a conservative result w.r.t. the theoretical parameter spanning. From the experimental point of view, 
the uncertainty on the measured central value makes the observation of a deviation on $g_{h\gamma\gamma}$ even more difficult.

\subsubsection{ILC}

The two dominant Higgs production reactions at future $e^+e^-$ Linear Colliders (LC) are the Higgs--strahlung mechanism and the $WW$ fusion process \cite{HiggsReviewI}. 
For an intermediate center-of-mass energy of $350 - 500$ GeV and an integrated luminosity of ${\cal L}=500$ fb$^{-1}$ (typically after one or two years of run), 
the experimental accuracy expected in the measurement of the $e^+e^- \to Zh$ cross section is $\pm 2.5\%$ for $m_h = 120$ GeV \cite{TeslaTDR}.
More recent studies dedicated to the International LC (ILC) performances \cite{ILC-FR} 
conclude that this accuracy can be improved down to $\pm 2\%$ with only $\sqrt{s}=250$ GeV and ${\cal L}=250$ fb$^{-1}$. For Higgs masses of $140-160$ GeV, the accuracy is a bit worst.
The cross section determination is independent of the Higgs decay modes: it is determined through a method analyzing the mass spectrum of the system recoiling against the $Z$ boson.
It is not the case for the process $e^+e^- \to h \nu \nu$ which is experimentally analyzed via the decay $h \to b \bar b$ and is measured with a weaker precision. We do not
consider this process here. The separation of the Higgs--strahlung mechanism and the $WW$ fusion process is partially controllable by properly choosing the beam polarization configurations. 
Let us consider RS points A and B as the results will be only slightly modified at $m_h = 150$ GeV due to the stronger EWPT limits on $M_{KK}$ and the weaker experimental sensitivity. 
Before presenting the numerical results, we mention that below $\sqrt{s}=500$ GeV the direct effect due to the exchange of KK $Z$ excitations in $e^+e^- \to Zh$ is negligible
compared to the $Z^0$--$Z^{KK}$ mixing effect on the $g_{hZZ}$ coupling. The RS corrections to the $Z$ boson mass, width and coupling to $e^+e^-$ are also negligible due to the strong
EWPT constraints.

Assuming the parameter set A/B, the ILC experiment would measure a cross section $\sigma(e^+e^- \to Zh)$ in the interval $[\sigma_-,\sigma_+]$ where, according to Eq.(\ref{intermed}),
\begin{eqnarray}
\sigma_\pm = \sigma^{\rm RS} \frac{1 \pm 2\delta\sigma/\sigma}{1 \mp 2\delta\sigma/\sigma} 
           = \sigma^{\rm SM} \frac{\sigma^{\rm RS}}{\sigma^{\rm SM}} \frac{1 \pm 2\delta\sigma/\sigma}{1 \mp 2\delta\sigma/\sigma},
\label{intermedILC}   
\end{eqnarray}
$\delta\sigma/\sigma$ being the experimental accuracy. Having replaced $\delta\sigma/\sigma$ by $2\delta\sigma/\sigma$, the probability to obtain a measure
in this interval is of $95\%$ like the $C.L.$ for this measurement. 
Taking $\sigma^{\rm RS} / \sigma^{\rm SM} = ( g^{\rm RS}_{hZZ} / g^{\rm SM}_{hZZ} )^2$ (SM radiative corrections are compensated) 
and the dependence $\delta\sigma/\sigma = \sqrt{\sigma^{\rm SM} / \sigma^{\rm RS}} \times 2 \%$ \cite{ILC-FR}, we find for the $[\sigma_-,\sigma_+]$ ranges:
$$
[0.28,0.37] \ \times \ \sigma^{\rm SM} \ \ \ \mbox{(point A), \ and,} \ \ \ [0.69,0.83] \ \times \ \sigma^{\rm SM} \ \ \ \mbox{(point B).}
$$
These intervals would translate into possible Higgs couplings measurements in the following ranges 
(to be compared with the starting theoretical values for $g^{\rm RS}_{hZZ} / g^{\rm SM}_{hZZ}$ in Table \ref{MainTable}):
$$
0.53 < \frac{g^{\rm exp}_{hZZ}}{g^{\rm SM}_{hZZ}} < 0.61 \ \ \ \mbox{(point A), \ and,} \ \ \ 0.83 < \frac{g^{\rm exp}_{hZZ}}{g^{\rm SM}_{hZZ}} < 0.91 \ \ \ \mbox{(point B).} 
$$
These precise `reconstructions' of the couplings illustrate the ILC capability of clearly discriminating between pure SM couplings and couplings affected by RS corrections, in the Higgs--gauge sector.
This clear distinction is explained by the possible large RS corrections combined with the high ILC sensitivity on the cross section. Note that for $m_h = 120$ GeV, the ILC
can test the whole range of possible $g^{\rm RS}_{hZZ}$ values when $g_{Z'}$ spans the entire allowed interval $0.72-1.57$ (point B - point A), assuming the smallest $M_{KK}$ value
allowed by EWPT. In other words, even the regions with small $g_{hZZ}$ deviations (case of point B) can be tested. These deviations would constitute indirect 
experimental signatures of the RS scenario.

Let us finish this part by making a comment on the possibility at ILC of discriminating between different models beyond the SM, using the Higgs--gauge sector. 
Under the hypothesis $m_h = 120$ GeV, the theoretical $g^{\rm RS}_{hZZ} / g^{\rm SM}_{hZZ}$ value cannot go below $0.57$ [see Table \ref{MainTable}] 
and the found experimental value at ILC could not be below $0.53$ (at $95\% C.L.$). Hence a measurement at the ILC of the coupling constant $g_{hZZ}$ completely below 
$0.53 \times g^{\rm SM}_{hZZ}$ would exclude the RS model. For instance in the Minimal Supersymmetric SM, such a coupling $g^{\rm MSSM}_{hZZ} / g^{\rm SM}_{hZZ} = \sin (\beta-\alpha) \leq 0.53$ 
is realizable for a CP--odd boson $A^0$ mass $m_A \lesssim 130$ GeV and $\tan \beta \simeq 30$ \cite{HiggsReviewII}.

\subsection{The Higgs boson pair production}

Let us assume for a while 
that the Higgs boson has been discovered and that its mass has been measured at LHC or ILC. Then an approach similar to the one adopted throughout this paper
could be followed. First, one would have access to the quartic Higgs self--coupling, denoted $\lambda_h$, via the measured Higgs mass,
\begin{eqnarray}
m_h^2 = 2 \ \lambda_h \tilde v^2 + \delta^{\rm SM} m^2_h ,
\label{HiggsMass}   
\end{eqnarray}
where $\delta^{\rm SM} m^2_h$ includes the SM quantum corrections. Then one would be able to deduce the triple Higgs coupling strength:
\begin{eqnarray}
g_{hhh} = 3 \ \lambda_h \tilde v + \delta^{\rm SM} g_{hhh} ,
\label{HiggsTriple}   
\end{eqnarray}
as well as the $hhVV$ coupling strength: 
\begin{eqnarray}
g_{hhVV} = \frac{g^2_V}{2} + \delta g_{hhVV} ,
\label{HiggsDouble}   
\end{eqnarray}
with $g^2_V=\{ g^2+g'^2 , g^2 \}$ and $\delta g_{hhVV}$ taking into account both SM radiative corrections and KK gauge mixing effects.

Therefore, the theoretical values for the rates of the (more challenging) Higgs boson pair production could be computed within the RS model. 
Indeed, this pair production proceeds 
e.g. at LHC through {\it (i)} the usual single Higgs production diagrams 
with the final Higgs leg being connected to two other Higgs fields via the $g_{hhh}$ coupling,
{\it (ii)} a `double' Higgs--strahlung process $qq\to hhV$ with two radiated Higgs bosons involving either 
two times the $g_{hVV}$ vertex [plus the $t$--channel contribution] or one time $g_{hhVV}$
and {\it (iii)} the Vector Boson Fusion (VBF) mechanism $qq\to V^*V^* qq\to hh qq$ with the $g_{hhVV}$ coupling instead of the $g_{hVV}$ one.
At ILC, the Higgs pair production occurs similarly trough {\it (i)} the usual Higgs production diagrams with the final Higgs connected 
to two Higgs fields via $g_{hhh}$, {\it (ii)} a double Higgs--strahlung reaction off $Z$ bosons $e^+e^-\to hhZ$ involving two times the $g_{hZZ}$ coupling
or one time $g_{hhZZ}$ and {\it (iii)} the VBF mechanism $e^+e^-\to V^*V^* \ell\ell \to hh \ell\ell$ [$\ell = e^\pm,\nu$] involving $g_{hhVV}$ couplings.

In conclusion, because of the theoretical relation between the Higgs mass and its couplings (via the $\lambda_h$, $\tilde v$ values), 
the $m_h$, $m_V$, $G_F$ measurements would permit to predict the $g^{\rm RS}_{hVV}$, $g^{\rm RS}_{hhVV}$, $g^{\rm RS}_{hhh}$ coupling constants and 
hence potentially to test the RS model via the Higgs pair production. This new test would be more difficult, in particular due to the smaller rates, 
but complementary to the single Higgs production at LHC or ILC.

\section{RS variants}
\label{variants}

In this last part, we discuss qualitatively the implications of the Higgs VEV corrections on the EWPT and Higgs couplings 
that we have treated in detail above, but within the other versions of warped extra dimension models.

First, an alternative scenario to the one we have considered is that ${\rm SU(2)_R}$ remains unbroken in the bulk ($\tilde M = 0$).
Then the dominant contribution to $T_{\rm RS}$ comes from the exchange of (excited) $t$ and $b$ quarks at the one--loop level. 
The estimation of this radiatively generated $T_{\rm RS}$ relies on a sum over fermion/boson KK towers which depends on the choice 
of quark representations. In such a case, $\tilde v$ would no more depend on $\tilde M$ but 
the correlation between $S_{\rm RS}$ and $T_{\rm RS}$ through would still be modified by the dependence 
of the Higgs VEV corrections on these parameters. Indeed $T_{\rm RS}$ would still depend on $\tilde v$, $M_{KK}$ and $g_{Z'}$.
Besides, the increase of $\tilde v$ w.r.t. $v_{SM}=245$ GeV would imply an enhancement of $S_{\rm RS}$ which tends to increase
the EWPT lower limit on $M_{KK}$, as we also find above.

Another possibility within the RS framework is to include large kinetic terms for the gauge fields on the IR brane
\cite{kinetics}, without bulk custodial symmetry. Such terms repel the KK mode wave functions from the brane so that
the KK gauge mixing effect, coming from the coupling of KK gauge fields to the Higgs boson located at the IR brane,
is reduced. Hence the Higgs VEV corrections are expected to be significantly weaker than here, 
and their effects on EWPT constraints as well as on Higgs couplings softer.

In the case of gauge--Higgs unification scenarios, where the extended bulk gauge symmetry generally contains the
custodial group ${\rm SU(2)_L\! \times\! SU(2)_R\! \times\! U(1)_X}$, the KK gauge mixing effect [induced by EWSB] 
should be of the same order as in the present paper since the Higgs profile is peaked on the TeV--brane 
(instead of being exactly confined as here). By consequence, comparable Higgs VEV corrections are expected
and in turn similar EWPT bounds on $M_{KK}$.

Finally, we mention the so--called gaugephobic Higgs models where the Higgs VEV on the brane can be much larger than here,
forcing then the lightest $W$ and $Z$ modes to move further from the brane \cite{Lillie,Caccia}.
In the limit of an infinite VEV, where the couplings between the Higgs and gauge bosons vanish,
one recovers the Higgsless gauge boundary conditions. In such models, to maintain compatibility
with EWPT, one must render the corrections to the $S$ parameter small. For that purpose, one has to allow all 
the light fermions to be spread in the bulk \cite{EWPTHless}. When their profile becomes approximately flat, 
their wave function being then orthogonal to the KK gauge boson ones, the contributions to $S$ can be made arbitrarily
small. Nevertheless, with such a fermion universality, 
one clearly looses the beauty of generating the mass hierarchy and flavor structure via
the simple geometrical mechanism of wave function overlapping.

\section{Conclusion}
\label{conclu}

Within the RS framework, the corrections to the Higgs boson VEV induced by the KK gauge mixing can be large. Those imply
an enhancement of the EWPT lower limit at $95.45 \% C.L.$ on $M_{KK}$ that can be larger than $+30 \%$ for $m_h=120$ GeV.  
Another important role of these RS corrections to the VEV is played in the calculation of the Higgs couplings. We find
that the Higgs couplings can be significantly reduced w.r.t. SM rendering the Higgs production detection at LHC more delicate. 
The Higgs rate suppressions also allow to pass the LEP2 constraint with $m_h \simeq 99$ GeV, a mass for which
the Higgs discovery at LHC would be tricky. Finally, the large deviations to the Higgs couplings due to extra dimensions provide 
an indirect way of testing the RS model: the LHC precision in light Higgs rate measurements would allow to explore KK boson
mass ranges above $4$ TeV (in agreement with EWPT) through Higgs production/decay channels involving the couplings $g_{hgg}$ 
and $g_{hVV}$. With the higher accuracies expected at ILC, even more clear signatures of the RS scenario may arise 
in the precise measurement of $g_{hZZ}$ deviations. 
\\
\\ 
\noindent \textbf{Acknowledgments:} 
The authors are grateful to H.~Bachacou, S.~Gopalakrishna, C.~Grojean, P.~Lutz, M.~Passera and F.~Richard 
for useful discussions. We also thank M.~Calvet for her contribution to the manuscript. 
This work is supported by the HEPTOOLS network and the A.N.R. {\it LFV-CPV-LHC} under project NT09\_508531.

\newpage

\appendix 
\noindent \textbf{\Large Appendix} \vspace{0.5cm} 

\renewcommand{\thesubsection}{A.\arabic{subsection}} 
\renewcommand{\theequation}{A.\arabic{equation}} 
\setcounter{subsection}{0} 
\setcounter{equation}{0} 

\section{Gauge boson mass matrices} 
\label{MassMat}

The neutral gauge boson mass matrix reads as (writing only the first KK mode contributions),
\begin{eqnarray}   
{\cal M}^2_0 =
\left (  
\begin{array}{ccc} 
g_Z^2\frac{\tilde v^2}{4} + \delta^{\rm SM}m^2_Z 
& g_Z^2\frac{\tilde v^2}{4} \sqrt{2 k \pi R_c} & - g_Z^2\frac{\tilde v^2}{4} \sqrt{2 k \pi R_c} \frac{g_{Z'}}{g_Z} \cos^2 \theta' \\  
g_Z^2\frac{\tilde v^2}{4} \sqrt{2 k \pi R_c} & M^2_{KK} + g_Z^2\frac{\tilde v^2}{4} (2 k \pi R_c) 
& - g_Z^2\frac{\tilde v^2}{4} (2 k \pi R_c) \frac{g_{Z'}}{g_Z} \cos^2 \theta' \\ 
- g_Z^2\frac{\tilde v^2}{4} \sqrt{2 k \pi R_c} \frac{g_{Z'}}{g_Z} \cos^2 \theta' 
& - g_Z^2\frac{\tilde v^2}{4} (2 k \pi R_c) \frac{g_{Z'}}{g_Z} \cos^2 \theta' 
& M'^2_{KK} + g_Z^2\frac{\tilde v^2}{4} (2 k \pi R_c) \frac{g_{Z'}^2}{g_Z^2} \cos^4 \theta'   
\end{array} 
\right ) ,
\nonumber\\
\label{NeutMassMat}   
\end{eqnarray}
with $ g_Z^2 = g^2+g'^2$.
The increasing factor $\sqrt{2 k \pi R_c}$ is the ratio of the $Z^{(1)}$ over $Z^0$ wave function amounts at the TeV--brane (where is stuck the Higgs boson). 
We have checked that numerically this ratio is not significantly different for the three first KK states (independently of the BC: $(++)$ or $(-+)$). 
However, the sign of the n{\it th} KK gauge wave function at the IR boundary goes like $(-1)^{n-1}$.
The three first KK masses are respectively $M_{KK}  \simeq (2.45;5.57;8.70) k e^{-\pi kR_{c}}$ and $M'_{KK} \simeq (2.40;5.52;8.65) k e^{-\pi kR_{c}}$.

The charged gauge boson mass matrix is (note the presence of the $\tilde M^2$ term):
\begin{eqnarray}   
{\cal M}^2_\pm =
\left (  
\begin{array}{ccc} 
g^2\frac{\tilde v^2}{4} + \delta^{\rm SM}m^2_W & g^2\frac{\tilde v^2}{4} \sqrt{2 k \pi R_c} & - g^2\frac{\tilde v^2}{4} \sqrt{2 k \pi R_c} \frac{\tilde g}{g} \\   
g^2\frac{\tilde v^2}{4} \sqrt{2 k \pi R_c} & M^2_{KK} + g^2\frac{\tilde v^2}{4} (2 k \pi R_c) 
& - g^2\frac{\tilde v^2}{4} (2 k \pi R_c) \frac{\tilde g}{g} \\ 
- g^2\frac{\tilde v^2}{4} \sqrt{2 k \pi R_c} \frac{\tilde g}{g}
& - g^2\frac{\tilde v^2}{4} (2 k \pi R_c) \frac{\tilde g}{g} 
& ( M'_{KK} + \frac{\tilde M^2}{4k} e^{-\pi kR_{c}} )^2 + g^2\frac{\tilde v^2}{4} (2 k \pi R_c) \frac{\tilde g^2}{g^2} 
\end{array} 
\right ) .
\nonumber\\
\label{ChargMassMat}   
\end{eqnarray}

\renewcommand{\thesubsection}{B.\arabic{subsection}} 
\renewcommand{\theequation}{B.\arabic{equation}} 
\setcounter{subsection}{0} 
\setcounter{equation}{0} 

\section{Gauge boson couplings} 
\label{CouplMat} 

The neutral gauge boson couplings to the Higgs boson are given by the following matrix (see Section \ref{couplings}),
\begin{eqnarray}   
{\cal C}_0 =
\left (  
\begin{array}{ccc} 
g_Z^2\frac{\tilde v^2}{4} + \delta^{\rm SM}g_{hZZ} 
& g_Z^2\frac{\tilde v^2}{4} \sqrt{2 k \pi R_c} & - g_Z^2\frac{\tilde v^2}{4} \sqrt{2 k \pi R_c} \frac{g_{Z'}}{g_Z} \cos^2 \theta' \\  
g_Z^2\frac{\tilde v^2}{4} \sqrt{2 k \pi R_c} & g_Z^2\frac{\tilde v^2}{4} (2 k \pi R_c) 
& - g_Z^2\frac{\tilde v^2}{4} (2 k \pi R_c) \frac{g_{Z'}}{g_Z} \cos^2 \theta' \\ 
- g_Z^2\frac{\tilde v^2}{4} \sqrt{2 k \pi R_c} \frac{g_{Z'}}{g_Z} \cos^2 \theta' 
& - g_Z^2\frac{\tilde v^2}{4} (2 k \pi R_c) \frac{g_{Z'}}{g_Z} \cos^2 \theta' 
& g_Z^2\frac{\tilde v^2}{4} (2 k \pi R_c) \frac{g_{Z'}^2}{g_Z^2} \cos^4 \theta'   
\end{array} 
\right ) 
\label{NeutCouplMat}   
\end{eqnarray}
where the irreducible quantum correction to the $hZZ$ vertex is,
\begin{equation} 
\delta^{\rm SM}g_{hZZ} = - \frac{3 (g^2+g'^2)}{16 \pi^2} \frac{m_b^2+m_t^2}{2} .
\label{DSMirrZ} 
\end{equation}
We do not include explicitly the SM radiative corrections due to the Higgs boson self--energy as those constitute 
a common factor in the RS and SM coupling $hZZ$, thus disappearing in their ratio that we will compute here.

The charged gauge boson couplings to the Higgs boson are alternatively determined by 
\begin{eqnarray}   
{\cal C}_\pm =
\left (  
\begin{array}{ccc} 
g^2\frac{\tilde v^2}{4} + \delta^{\rm SM}g_{hWW} & g^2\frac{\tilde v^2}{4} \sqrt{2 k \pi R_c} & - g^2\frac{\tilde v^2}{4} \sqrt{2 k \pi R_c} \frac{\tilde g}{g} \\   
g^2\frac{\tilde v^2}{4} \sqrt{2 k \pi R_c} & g^2\frac{\tilde v^2}{4} (2 k \pi R_c) 
& - g^2\frac{\tilde v^2}{4} (2 k \pi R_c) \frac{\tilde g}{g} \\ 
- g^2\frac{\tilde v^2}{4} \sqrt{2 k \pi R_c} \frac{\tilde g}{g}
& - g^2\frac{\tilde v^2}{4} (2 k \pi R_c) \frac{\tilde g}{g} 
& g^2\frac{\tilde v^2}{4} (2 k \pi R_c) \frac{\tilde g^2}{g^2} 
\end{array} 
\right ) 
\label{ChargCouplMat}   
\end{eqnarray}
the irreducible quantum correction to the $hWW$ vertex being: 
\begin{equation} 
\delta^{\rm SM}g_{hWW} = - \frac{3 g^2}{16 \pi^2} \bigg ( \frac{m_b^2+m_t^2}{4} - \frac{m_b^4 \ ln(m_b^2 / m_t^2)}{2(m_t^2-m_b^2)} \bigg ) .
\label{DSMirrW} 
\end{equation}

\renewcommand{\thesubsection}{C.\arabic{subsection}} 
\renewcommand{\theequation}{C.\arabic{equation}} 
\setcounter{subsection}{0} 
\setcounter{equation}{0}

\section{Bottom and top quark mass matrices} 
\label{FermMat}

In the field basis $\Psi^t_L \equiv (b^{(0)}_L, b^{(1)}_L, b^{c(1)}_L, b''^{(1)}_L, b'^{c(1)}_L)^t$, 
$\Psi^t_R \equiv (b^{c(0)}_R, b^{(1)}_R, b^{c(1)}_R, b''^{(1)}_R, b'^{c(1)}_R)^t$, 
the 4D bottom quark mass matrix, up to the first KK modes 
(numerical analysis includes first and second KK modes) for our model is,
\begin{eqnarray}   
{\cal M}_b =
\left (  
\begin{array}{ccccc}
\epsilon_b \tilde{v}_b c_{\theta} f^{(0)*}_{c_2} f^{(0)}_{c_{b_R}} & 0 & \tilde{v}_b c_{\theta} f^{(0)*}_{c_2} f^{(1)}_{c_{b_R}} & 0 & \sqrt{2} \tilde{v}_t s_{\theta} f^{(0)*}_{c_1} g^{(1)}_{c_{t_R}} \\
\tilde{v}_b c_{\theta} f^{(1)*}_{c_2} f^{(0)}_{c_{b_R}} & s^2_{\theta} m^{(1)}_{c_1} + c^2_{\theta} m^{(1)}_{c_2} & \tilde{v}_b c_{\theta} f^{(1)*}_{c_2} f^{(1)}_{c_{b_R}} & 0 & \sqrt{2} \tilde{v}_t s_{\theta} f^{(1)*}_{c_1} g^{(1)}_{c_{t_R}} \\
0 & 0 & m^{(1)}_{c_{b_R}} & 0 & 0 \\
{1\over\sqrt{2}} \tilde{v}_b g^{(1)*}_{c_2} f^{(0)}_{c_{b_R}} & 0 & {1\over\sqrt{2}} \tilde{v}_b c_{\theta} g^{(1)*}_{c_2} f^{(1)}_{c_{b_R}} & m'^{(1)}_{c_2} & 0 \\
0 & 0 & 0 & 0 & m'^{(1)}_{c_{t_R}}
\end{array}
\right ) .
\label{BotMassMat}
\end{eqnarray}

In the field basis $\Phi^t_L \equiv (t^{(0)}_L, t^{(1)}_L, t^{c(1)}_L, t'^{(1)}_L)^t$, 
$\Phi^t_R \equiv (t^{c(0)}_R, t^{(1)}_R, t^{c(1)}_R, t'^{(1)}_R)^t$, 
the top quark mass matrix up to the first KK states (same comment for the numerical analysis) is given by:
\begin{eqnarray}   
{\cal M}_t =
\left (  
\begin{array}{cccc}
\epsilon_t \tilde{v}_t s_{\theta} f^{(0)*}_{c_1} f^{(0)}_{c_{t_R}} & 0 & \tilde{v}_t s_{\theta} f^{(0)*}_{c_1} f^{(1)}_{c_{t_R}} & 0 \\
\tilde{v}_t s_{\theta} f^{(1)*}_{c_1} f^{(0)}_{c_{t_R}} & s^2_{\theta} m^{(1)}_{c_1} + c^2_{\theta} m^{(1)}_{c_2} & \tilde{v}_t s_{\theta} f^{(1)*}_{c_1} f^{(1)}_{c_{t_R}} & 0 \\
0 & 0 & m^{(1)}_{c_{t_R}} & 0 \\
\tilde{v}_t s_{\theta} g^{(1)*}_{c_1} f^{(0)}_{c_{t_R}} & 0 & \tilde{v}_t s_{\theta} g^{(1)*}_{c_1} f^{(1)}_{c_{t_R}} & m'^{(1)}_{c_1}
\end{array}
\right ) .
\label{TopMassMat}
\end{eqnarray}

$\epsilon_{b,t}$ is described in Section \ref{couplYuk}.
Besides, in our notations,  
$\tilde{v}_{b,t} = \lambda^{5D}_{b,t} \tilde{v}/\sqrt{2}kR_c$, $m^{(n)}_c$ ($m'^{(n)}_c$) is the $n$--{\it th} KK mass 
for $(++)$ ($(-+)$) BC fields, $c_{\theta}=\cos \theta$ ($s_{\theta}=\sin \theta$) and
$\theta$ is the effective angle of the mixing between the two left multiplets. 
$f^{(n)}_c$ and $g^{(n)}_c$ are respectively the fermion wave functions along the $5$--{\it th} dimension with $(++)$ and $(-+)$ BC, 
whose values are taken at the position of the TeV--brane, $x_5 = \pi R_c$ (where the Higgs boson is confined). For instance 
(see e.g. \cite{AgaSer} for excited profiles):
\begin{eqnarray}
f^{(0)}_c (x_5) \equiv \sqrt{{(1-2c)k R_c}\over{e^{(1-2c)\pi k R_c}-1}}e^{({1\over2}-c)k x_5}.
\end{eqnarray} 
The zeroes in the bottom mass matrix originate from the fact that the fields 
$b^{c(1)}_L$, $b'^{c(1)}_L$, $b^{(1)}_R$ and $b''^{(1)}_R$ (with $n=1,2$) have Dirichlet BC on the TeV--brane 
and thus, do not couple to the Higgs boson. For the top mass matrix, it originates from the fact that 
$t^{c(1)}_L$, $t^{(1)}_R$ and $t'^{(1)}_R$ have Dirichlet BC on the TeV-brane.

\renewcommand{\thesubsection}{D.\arabic{subsection}} 
\renewcommand{\theequation}{D.\arabic{equation}} 
\setcounter{subsection}{0} 
\setcounter{equation}{0}

\section{Bottom and top quark Coupling Matrices}
\label{FermCoupl}

The bottom quark Yukawa coupling matrix reads as,
\begin{eqnarray}   
{\cal C}_b =
\left (  
\begin{array}{ccccc}
\epsilon_b \tilde{v}_b c_{\theta} f^{(0)*}_{c_2} f^{(0)}_{c_{b_R}} & 0 & \tilde{v}_b c_{\theta} f^{(0)*}_{c_2} f^{(1)}_{c_{b_R}} & 0 & \sqrt{2} \tilde{v}_t s_{\theta} f^{(0)*}_{c_1} g^{(1)}_{c_{t_R}} \\
\tilde{v}_b c_{\theta} f^{(1)*}_{c_2} f^{(0)}_{c_{b_R}} & 0 & \tilde{v}_b c_{\theta} f^{(1)*}_{c_2} f^{(1)}_{c_{b_R}} & 0 & \sqrt{2} \tilde{v}_t s_{\theta} f^{(1)*}_{c_1} g^{(1)}_{c_{t_R}} \\
0 & 0 & 0 & 0 & 0 \\
{1\over\sqrt{2}} \tilde{v}_b g^{(1)*}_{c_2} f^{(0)}_{c_{b_R}} & 0 & {1\over\sqrt{2}} \tilde{v}_b c_{\theta} g^{(1)*}_{c_2} f^{(1)}_{c_{b_R}} & 0 & 0 \\
0 & 0 & 0 & 0 & 0
\end{array}
\right ) ,
\label{CBotMassMat}
\end{eqnarray}
It is obtained from the bottom mass matrix defined in previous appendix, but with the KK mass terms set to zero. 
Hence, it can be defined through the relation:
\begin{eqnarray} 
{\cal C}_b \equiv \tilde{v}{{\partial\mathcal{M}_b}\over{\partial \tilde{v}}} , 
\label{exPartial}
\end{eqnarray}
as may be useful.

The dimensionful top quark Yukawa couplings can be derived from the matrix:
\begin{eqnarray}   
{\cal C}_t =
\left (  
\begin{array}{cccc}
\epsilon_t \tilde{v}_t s_{\theta} f^{(0)*}_{c_1} f^{(0)}_{c_{t_R}} & 0 & \tilde{v}_t s_{\theta} f^{(0)*}_{c_1} f^{(1)}_{c_{t_R}} & 0 \\
\tilde{v}_t s_{\theta} f^{(1)*}_{c_1} f^{(0)}_{c_{t_R}} & 0 & \tilde{v}_t s_{\theta} f^{(1)*}_{c_1} f^{(1)}_{c_{t_R}} & 0 \\
0 & 0 & 0 & 0 \\
\tilde{v}_t s_{\theta} g^{(1)*}_{c_1} f^{(0)}_{c_{t_R}} & 0 & \tilde{v}_t s_{\theta} g^{(1)*}_{c_1} f^{(1)}_{c_{t_R}} & 0
\end{array}
\right ) .
\label{CTopMassMat}
\end{eqnarray}

\renewcommand{\thesubsection}{E.\arabic{subsection}} 
\renewcommand{\theequation}{E.\arabic{equation}} 
\setcounter{subsection}{0} 
\setcounter{equation}{0} 

\section{Fermion Sum Rule}
\label{FermionSumRule}

In this Appendix, we demonstrate analytically some useful theoretical relations which allow to take into account the full
KK quark tower in the calculation of the gluon--gluon fusion mechanism amplitude, within the RS scenario. 
This generalizes results obtained in the case of a bulk Higgs boson within the framework of gauge--Higgs unification \cite{Adam}.
These relations also allow to implement the full KK charged fermion tower when calculating the loop-induced
$h\gamma\gamma$ coupling.

For that, we use the same conventions/notations as in the Section \ref{couplYuk}:
$\mathcal{M}_f$ is defined as the mass matrix of a particular fermion $\Psi_{L/R}$ in the interaction basis. 
We have obtained $\mathcal{C}'_f/\tilde v$ to be the Yukawa matrix in the mass eigenbasis,
where $\mathcal{C}'_f=U_L\mathcal{C}_f U^{\dagger}_R$; $\mathcal{C}_f$ being the same matrix as $\mathcal{M}_f$ 
but with the KK masses set to zero.

The matrices $\mathcal{M}_f$ and $\mathcal{C}'_f$ are thus linked through the relation 
(see for instance Eq.(\ref{exPartial})):
\begin{eqnarray}
\mathcal{C}'_f \equiv U_L . \tilde v{{\partial\mathcal{M}_f}\over{\partial \tilde v}} . U^{\dagger}_R
\end{eqnarray}

Let us define here, for simplicity about the subscript notation, 
$m_i \equiv \mathcal{M}'_{f}|_{ii}$ and $\lambda^{\rm RS}_i \equiv (\mathcal{C}'_f|_{ii})/\tilde v$ for a given fermion
within RS. Then we have
\begin{eqnarray}
\sum_i{{\lambda^{\rm RS}_i\,\tilde v}\over{m_i}} \equiv \mbox{Tr}\left(\mathcal{M'}_{f}^{-1}.{\mathcal{C'}_{f}}\right) 
\quad\mbox{where}\quad \mathcal{M'}_{f}^{-1}.\mathcal{M'}_{f} = 1.
\end{eqnarray}
and can rewrite this trace as (accordingly to Eq.(\ref{MfDIAGO})):
\begin{eqnarray}
\sum_i{{\lambda^{\rm RS}_i\,\tilde v}\over{m_i}}= \mbox{Tr}\left( U_R \mathcal{M}_f^{-1} U^{\dagger}_L.
U_L \tilde v{\partial\mathcal{M}_f\over{\partial \tilde v}} U^{\dagger}_R \right) =
\mbox{Tr}\left( \tilde v{\partial\mathcal{M}_f\over{\partial \tilde v}}.\mathcal{M}_f^{-1} \right)
\end{eqnarray}
\begin{eqnarray}
\sum_i{{\lambda^{\rm RS}_i\,\tilde v}\over{m_i}} = \tilde v{\partial\over{\partial \tilde v}} \mbox{Tr}\left( ln \ \mathcal{M}_f \right) 
= \tilde v{\partial\over{\partial \tilde v}} \ ln \left( \mbox{Det}\mathcal{M}_f  \right)
\end{eqnarray}

Applying this result to SM fermion mass matrices such as the ones in the Appendix \ref{FermMat}
(including even possibly the entire KK tower contribution),
\begin{eqnarray}
\sum_i{{\lambda^{\rm RS}_i\,\tilde v}\over{m_i}} = \tilde v{\partial\over{\partial \tilde v}} \sum_i \ ln \ \mathcal{M}_f|_{ii} 
= \tilde v{\partial\over{\partial \tilde v}} \ ln \ \mathcal{M}_f|_{11} = 1
\label{eq:FermRule}
\end{eqnarray}
and for exotic fermions without zero--mode, as for instance the $q_{(-7/3)}'$ field appearing in a multiplet of Eq.(\ref{Model}),
\begin{eqnarray}
\sum_i{{\lambda^{\rm RS}_i\,\tilde v}\over{m_i}} = \tilde v{\partial\over{\partial \tilde v}} \sum_i \ ln \ \mathcal{M}_f|_{ii} = 0 .
\label{eq:FermRulePrime}
\end{eqnarray}

\vskip .5cm
\underline{Application to the KK fermionic contributions of the effective $hgg$ and $h\gamma\gamma$ couplings:}
\\ \\
In both Eq.(\ref{eq:hgg}) and Eq.(\ref{eq:hgaga}), the fermionic contribution to the amplitude reads, for each independent KK tower, 
$\sum_{\{f\}} {{\lambda^{\rm RS}_f\tilde{v}}\over{m_f}} A^h_{1/2}(\tau_f)$, up to irrelevant global color and electric charge factors. 
The sum over $\{f\}$ denotes the sum for a corresponding fermion $f$, of its zero--mode and all KK excitations it couples to 
through the Higgs field.

Using the properties of the spin--$1/2$ form factor, one can set that for all KK excitations, $A^h_{1/2}(\tau_{f^{KK}}) \simeq 1$ 
with a high precision (up to 1 per 1000). Having so, one can separate the contribution in the tower of the first fermion eigenstate
(generally mainly composed by the zero--mode component) noted $f^0$ from the heavier eigenstates (mainly made of KK modes)
noted $f^{KK}$'s for simplicity here:
\begin{eqnarray}
&&\sum_{\{f\}} {{\lambda^{\rm RS}_f\tilde{v}}\over{m_f}} A^h_{1/2}(\tau_f) 
\equiv {{\lambda^{\rm RS}_{f^0}\tilde{v}}\over{m_{f^0}}} A^h_{1/2}(\tau_{f^0}) 
+ \sum_{KK} {{\lambda^{\rm RS}_{f^{KK}}\tilde{v}}\over{m_{f^{KK}}}} A^h_{1/2}(\tau_{f^{KK}}) 
= {{\lambda^{\rm RS}_{f^0}\tilde{v}}\over{m_{f^0}}} A^h_{1/2}(\tau_{f^0}) 
+ \sum_{KK} {{\lambda^{\rm RS}_{f^{KK}}\tilde{v}}\over{m_{f^{KK}}}}  \nonumber\\
&&\sum_{\{f\}} {{\lambda^{\rm RS}_f\tilde{v}}\over{m_f}} A^h_{1/2}(\tau_f) 
= {{\lambda^{\rm RS}_{f^0}\tilde{v}}\over{m_{f^0}}} A^h_{1/2}(\tau_{f^0}) 
+ \left( 1 - {{\lambda^{\rm RS}_{f^0}\tilde{v}}\over{m_{f^0}}} \right) 
= 1 + {{\lambda^{\rm RS}_{f^0}\tilde{v}}\over{m_{f^0}}} \left( A^h_{1/2}(\tau_{f^0}) - 1 \right) ,
\label{SUMq}
\end{eqnarray}
where we have used the relation (\ref{eq:FermRule}). 
It is a remarkable feature that the KK sum can be reflected in a few properties from the lightest mode.

Note that for an exotic fermion $f'$, with no zero--mode, the sum even cancels:
\begin{eqnarray}
\sum_{\{f'\}} {{\lambda^{\rm RS}_{f'}\tilde{v}}\over{m_{f'}}} A^h_{1/2}(\tau_{f'}) 
\equiv \sum_{KK} {{\lambda^{\rm RS}_{f^{KK}}\tilde{v}}\over{m_{f^{KK}}}} A^h_{1/2}(\tau_{f^{KK}}) 
= \sum_{KK} {{\lambda^{\rm RS}_{f^{KK}}\tilde{v}}\over{m_{f^{KK}}}} = 0
\label{SUMqprime}
\end{eqnarray}
according to the relation (\ref{eq:FermRulePrime}).

\renewcommand{\thesubsection}{F.\arabic{subsection}} 
\renewcommand{\theequation}{F.\arabic{equation}} 
\setcounter{subsection}{0} 
\setcounter{equation}{0} 

\section{Boson Sum Rule}
\label{BosonSumRule}

Here are shown some theoretical sum rules for the bosons, similarly to the fermion case of previous appendix,
but which apply now to the computation of the effective $h\gamma\gamma$ vertex. 

The coupling matrix ${\cal C}'_{\pm}$ and mass matrix ${\cal M}^2_{\pm}$ for the charged gauge bosons are linked via,
\begin{eqnarray}
{\cal C}'_{\pm} \equiv V . \tilde v^2{{\partial{\cal M}^2_{\pm}}\over{\partial \tilde v^2}} . V^{\dagger}
\end{eqnarray}

Reminding $m_{W^n}^2 = \mathcal{M'}^2_\pm|_{nn}$ and $g^{\rm RS}_{hW^nW^n} = 2 (\mathcal{C'}_\pm|_{nn})/\tilde v$
from Section \ref{hVVsec},
the loop part from the gauge boson KK tower can then be written as [see Eq.(\ref{McEIGEN})]:
\begin{eqnarray}
&&\sum_{n} {{g^{\rm RS}_{W^n} \tilde v}\over{2m_{W^n}^2}}
= \sum_{n} {{\mathcal{C'}_\pm|_{nn}}\over{\mathcal{M'}^2_\pm|_{nn}}} 
= \mbox{Tr}\left([\mathcal{M'}^2_{\pm}]^{-1}.{\mathcal{C'}_{\pm}}\right) 
= \mbox{Tr}\left(V [\mathcal{M}^2_{\pm}]^{-1} V^{\dagger}.V 
           \tilde v^2{{\partial\mathcal{M}^2_{\pm}}\over{\partial \tilde v^2}} V^{\dagger}\right) 
 \nonumber\\
&&\sum_{n} {{g^{\rm RS}_{W^n} \tilde v}\over{2m_{W^n}^2}}
= \mbox{Tr}\left(\tilde v^2{{\partial\mathcal{M}^2_{\pm}}\over{\partial \tilde v^2}}.[\mathcal{M}^2_{\pm}]^{-1}\right)
= \tilde v^2{{\partial}\over{\partial \tilde v^2}}\mbox{Tr}\left(ln \ \mathcal{M}^2_{\pm}\right) 
= \tilde v^2{{\partial}\over{\partial \tilde v^2}} \ ln \left( \mbox{Det}\mathcal{M}^2_{\pm}  \right) .
\end{eqnarray}
Obviously, this relation also holds for the neutral gauge boson KK tower, and one can check that 
\begin{eqnarray}
\mbox{Det}\mathcal{M}^2_{\pm} = g^2{{\tilde{v}^2}\over{4}} \prod_{n} \ M_{KK}^{(n)2} 
\ \bar M_{KK}^{\prime (n)2},
\end{eqnarray}
where $\bar M_{KK}^{\prime(n)2} = ( M_{KK}^{\prime(n)} + \frac{\tilde M^2}{4k} e^{-\pi kR_{c}} )^2$,
so that, at the end:
\begin{eqnarray}
\sum_{n} {{g^{\rm RS}_{W^n} \tilde v}\over{2m_{W^n}^2}} 
= \tilde{v}^2{\partial\over{\partial \tilde{v}^2}} \ ln (g^2{{\tilde{v}^2}\over{4}}) = 1 .
\end{eqnarray}

\vskip .5cm
\underline{Application to the KK $W$ contributions of the effective $h\gamma\gamma$ coupling:}
\\ \\
Applying the above result to the EW gauge boson KK tower contribution 
in the $h\gamma\gamma$ effective coupling allows one to rewrite:
\begin{eqnarray}
&&\sum_{n} {{\mathcal{C'}_\pm|_{nn}}\over{\mathcal{M'}^2_\pm|_{nn}}} A^h_{1}(\tau_{W^{n}}) \
\equiv {{g^{\rm RS}_{_{hWW}}\tilde{v}}\over{2m_W^2}} A^h_{1}(\tau_W) 
+ \sum_{n\geq1} {{\mathcal{C'}_\pm|_{nn}}\over{\mathcal{M'}^2_\pm|_{nn}}} A^h_{1}(\tau_{W^{n}}) \nonumber\\
&&\sum_{n} {{\mathcal{C'}_\pm|_{nn}}\over{\mathcal{M'}^2_\pm|_{nn}}} A^h_{1}(\tau_{W^{n}}) 
= {{g^{\rm RS}_{_{hWW}}\tilde{v}}\over{2m_W^2}} A^h_{1}(\tau_W) - 7 \sum_{n\geq1} {{\mathcal{C'}_\pm|_{nn}}\over{\mathcal{M'}^2_\pm|_{nn}}} 
= {{g^{\rm RS}_{_{hWW}}\tilde{v}}\over{2m_W^2}} A^h_{1}(\tau_W) - 7 \left( 1 - {{g^{\rm RS}_{_{hWW}}\tilde{v}}\over{2m_W^2}} \right) \nonumber\\
&&\sum_{n} {{\mathcal{C'}_\pm|_{nn}}\over{\mathcal{M'}^2_\pm|_{nn}}} A^h_{1}(\tau_{W^{n}}) 
= - 7 + {{g^{\rm RS}_{_{hWW}}\tilde{v}}\over{2m_W^2}} \left( 7 + A^h_{1}(\tau_W) \right) ,
\end{eqnarray}
where we have used the properties of the spin--$1$ form factor, so one can set that for all KK excitations, 
$A^h_{1}(\tau_{W^{n}}) = -7$ again at a high accuracy level (up to 1 per 1000).

\renewcommand{\thesubsection}{G.\arabic{subsection}} 
\renewcommand{\theequation}{G.\arabic{equation}} 
\setcounter{subsection}{0} 
\setcounter{equation}{0}

\section{Experimental sensitivity on Higgs observables} 
\label{Consider}

Given the expected relative precision $\delta{\cal Q}/{\cal Q}$ on a quantity ${\cal Q}$ to be measured in the Higgs boson sector, 
one can predict that there is a probability of $68\%$ that the measurement will be at most given by,
\begin{eqnarray}
{\cal Q}_{\rm exp}^{\rm max} = {\cal Q}_{\rm RS} + \frac{\delta{\cal Q}}{{\cal Q}} {\cal Q}_{\rm exp}^{\rm max} ,
\label{DQsQ}   
\end{eqnarray}
assuming that the exact value ${\cal Q}_{\rm RS}$ is the one obtained theoretically in the RS model. 
Then an experimental error of $(\delta{\cal Q}/{\cal Q}){\cal Q}_{\rm exp}^{\rm max}$ would be taken.
Hence, the largest experimental value that can be expected at a $68 \% C.L.$ is
\begin{eqnarray}
{\cal Q}_{\rm exp}^{\rm limit} = {\cal Q}_{\rm RS} + \ 2 \times \ \frac{\delta{\cal Q}}{{\cal Q}} {\cal Q}_{\rm exp}^{\rm max} .
\label{Qup}   
\end{eqnarray}
From Eq.(\ref{DQsQ}) and Eq.(\ref{Qup}), we deduce
\begin{eqnarray}
{\cal Q}_{\rm exp}^{\rm limit} = {\cal Q}_{\rm RS} + \ 2 \times \ \frac{\delta{\cal Q}}{{\cal Q}} \frac{{\cal Q}_{\rm RS}}{1-\delta{\cal Q}/{\cal Q}} 
= {\cal Q}_{\rm RS} \frac{1+\delta{\cal Q}/{\cal Q}}{1-\delta{\cal Q}/{\cal Q}}.
\label{intermed}   
\end{eqnarray}
The condition to observe experimentally an RS negative correction to ${\cal Q}$, or in other words to measure a ${\cal Q}$ value smaller than the SM one: ${\cal Q}_{\rm SM}$, reads as
\begin{eqnarray}
{\cal Q}_{\rm SM} > {\cal Q}_{\rm exp}^{\rm limit} .
\label{condition}   
\end{eqnarray}
Assuming $\delta{\cal Q}/{\cal Q}<1$, one obtains the final condition from combining Eq.(\ref{intermed})-(\ref{condition}),
\begin{eqnarray}
\frac{{\cal Q}_{\rm RS}}{{\cal Q}_{\rm SM}} < \frac{1-\delta{\cal Q}/{\cal Q}}{1+\delta{\cal Q}/{\cal Q}} .
\label{final}   
\end{eqnarray}


\end{document}